\documentclass[useAMS,usenatbib,usegraphicx]{mn2e}

\usepackage{amssymb}
\newcommand{\msano}{{\rm M}_\odot ~{\rm yr}^{-1}}
\newcommand{\peg}{V374~Peg}
\newcommand{\alf}{Alfv\'en}

\title{Powerful Winds from Low-Mass Stars: \peg}
\author[A.~A.~Vidotto et al.]{A.~A.~Vidotto$^{1}$\thanks{E-mail: Aline.Vidotto@st-andrews.ac.uk}, {M.~Jardine}$^{1}$, {M.~Opher}$^{2}$, {J.~F.~Donati}$^{3}$ and {T.~I.~Gombosi}$^{4}$\\
$^{1}$SUPA, School of Physics and Astronomy, University of St Andrews, North Haugh, St Andrews, KY16 9SS, UK\\
$^{2}$George Mason University, 4400 University Drive, Fairfax, VA, 22030-4444, USA\\
$^{3}$LATT - CNRS/Universit\'e de Toulouse, 14 Av.~E.~Belin, Toulouse, F-31400, France\\
$^{4}$University of Michigan, 1517 Space Research Building, Ann Arbor, MI, 48109-2143, USA}
\begin{document}

\date{Accepted . Received ; in original form }

\pagerange{\pageref{firstpage}--\pageref{lastpage}} \pubyear{2010}

\maketitle

\label{firstpage}

%%%%%%%%%%%%%%%%%%%%%%%%%%%%%%%%%%%%%%%%%%%%%%%%%%%%%%%%%%%%%%%%%%%%%%%%
\begin{abstract}
The M dwarf \peg\ (M4) is believed to lie near the theoretical mass threshold for fully convective interiors. Its rapid rotation ($P=0.44$~days) along with its intense magnetic field point toward magneto-centrifugal acceleration of a coronal wind. In this work, we investigate the structure of the coronal wind of \peg\ by means of three-dimensional magnetohydrodynamical (MHD) numerical simulations. For the first time, an observationally derived surface magnetic field map is implemented in MHD models of stellar winds for a low mass star. By self-consistently taking into consideration the interaction of the outflowing wind with the magnetic field and vice versa, we show that the wind of \peg\ deviates greatly from a low-velocity, low-mass-loss rate solar-type wind. We have found general scaling relations for the terminal velocities, mass-loss rates, and spin-down times of highly magnetized M dwarfs. In particular, for \peg , our models show that terminal velocities across a range of stellar latitudes reach $u_\infty \simeq (1500$ -- $2300) n_{12}^{-1/2}~{\rm km~s}^{-1}$, where $n_{12}$ is the coronal wind base density in units of $10^{12}~{\rm cm}^{-3}$, while the mass-loss rates are about $\dot{M} \simeq 4 \times 10^{-10} n_{12}^{1/2}~\msano$. We also evaluate the angular-momentum loss of \peg, which presents a rotational braking timescale $\tau \simeq 28 {n_{12}^{-1/2}}$~Myr. Compared to observationally derived values from period distributions of stars in open clusters, this suggests that \peg\ may have low coronal base densities $(\lesssim 10^{11}~{\rm cm}^{-3})$. We show that the wind ram pressure of \peg\ is about $5$ orders of magnitude larger than for the solar wind. Never the less, a small planetary magnetic field intensity ($\sim 0.1$~G) is able to shield a planet orbiting at $1~$AU against the erosive effects of the stellar wind. However, planets orbiting inside the habitable zone of \peg, where the wind ram pressure is higher, might be facing a more significant atmospheric erosion. In that case, higher planetary magnetic fields of, at least, about half the magnetic field intensity of Jupiter, are required to protect the planet's atmosphere.
\end{abstract}

%%%%%%%%%%%%%%%%%%%%%%%%%%%%%%%%%%%%%%%%%%%%%%%%%%%%%%%%%%%%%%%%%%%%%%%%
\section{INTRODUCTION}
M dwarf (dM) stars comprise the most abundant group of stars in the Galaxy. Due to their small masses \citep[$\sim0.06$ -- $0.8$~M$_\odot$,][]{2007AsBio...7...85S}, they have long main-sequence lifetimes, which exceed the Hubble time. They are low-luminosity objects and because of that, their habitable zone, where liquid water may be found, is believed to lie very close to the star \citep{1993Icar..101..108K}. This makes dM stars prime targets in searches for terrestrial habitable planets. 

From the point of view of stellar evolutionary theory, dM stars lie on the boundary between fully convective stars (spectral types later than $\sim$M4) and stars with a radiative core (spectral types earlier than $\sim$M4). The latter are believed to host magnetic fields generated by solar-type dynamos, where the layer between the radiative core and the convective outer shell (the tachocline) produces rotational shear, which, in conjunction with convective motions, is able to amplify and maintain surface magnetic fields. The generation of magnetic fields in fully convective stars is still a matter of debate though. Because such stars lack a tachocline, the dynamo mechanism should be different from a solar-type dynamo. Also, their nearly solid-body rotation indicates that the magnetic fields should cause a quenching of the surface differential rotation, an ingredient believed to be essential in field generation. 

Although details of the dynamo mechanism may still be unknown, the change in the stellar internal conditions appears to be reflect in the shift in the magnetic field large-scale topology around spectral type M4. The observation of surface magnetic field distributions suggests that early-dM stars host weak large-scale fields with dominantly toroidal and non-axisymmetric poloidal configurations \citep{2008MNRAS.390..545D}, while mid-dM stars host strong, mainly axi-symmetric large-scale poloidal fields \citep{2008MNRAS.390..567M}. 

This topology shift should produce a corresponding change in the coronal structure of the star, and, in particular, in coronal tracers such as X-ray and radio emissions. However, both the X-ray luminosity $L_X$ as well as the quiescent radio luminosity $L_{\rm rad}$ do not present sharp transitions across the M4 boundary. $L_{\rm rad}$, for instance, remains relatively the same for all the spectral subtypes of dM stars, while $L_X$ is roughly proportional to the bolometric luminosity $L_{\rm bol}$, presenting a saturation limit of $L_X/L_{\rm bol} \sim 10^{-3}$ out to spectral types M7 \citep{1998A&A...331..581D, 2010ApJ...709..332B}. As a consequence, a correlation between radio and X-ray emissions is observed for almost all the spectral subtypes of dM stars, with a sharp transition being seen only around spectral type M7 \citep{1993ApJ...415..236G, 2006ApJ...648..629B, 2010ApJ...709..332B}. Because of such a correlation, it is often suggested that the origins of both X-ray emission, which is believed to trace the hot plasma contained in magnetic loop structures, and radio emission, which is believed to be produced by gyro-synchrotron radiation, are common and most probably due to the coronal magnetic fields. 

Regarding rotation, the evolution of dM stars can be inferred from observations of open clusters at different ages \citep{2006MNRAS.370..954I, 2007MNRAS.381.1638S, 2009ApJ...691..342H, 2009MNRAS.400..451, 2009ApJ...695..679M}. In young ($\lesssim 700~$Myr) open clusters, dM stars still present high rotation rates, which suggests that angular momentum losses at the early main-sequence phase are negligible for them \citep{2009IAUS..258..363I}. However, as the cluster ages ($\gtrsim 700~$Myr), the number of rapidly rotating dM stars decreases, implying that there should exist a mechanism of angular momentum removal that acts on time-scales of a few hundred million years \citep{2007MNRAS.381.1638S}. For solar-like main sequence stars, the magnetised stellar wind is believed to spin down the star by carrying away stellar angular momentum. It has been observationally established that the angular velocity rate $\Omega_0$ for solar-like stars varies as a function of age $t$ as $\Omega_0 \propto t^{-1/2}$ \citep{1972ApJ...171..565S}. However, it seems that the empirical \citeauthor{1972ApJ...171..565S}'s law is not valid for low-mass stars, suggesting that a solar-type wind (i.e., with low velocities and mass-loss rates) cannot reproduce the rotational evolution of fully-convective stars. 

The existence of hot coronae, rapid rotation, and high levels of magnetic activity in dM stars suggests the presence of winds with an enhanced mass loss as compared to the solar wind. However, the low-density, optically thin winds of these stars prevents the observation of traditional mass-loss signatures, such as P Cygni profiles. The still unobserved high mass-loss rates from dM stars could be able to disperse debris discs, explaining why discs around dM stars older than $\gtrsim 10$~Myr are scarcely found \citep{2005ApJ...631.1161P}. 

Estimates of mass-loss rates from dM stars vary considerably. It has been suggested that the coronal winds of dM stars, despite of being very tenuous, possess mass-loss rates ($\dot{M}$) that can considerably exceed the solar value ($\dot{M}_\odot \simeq 2 \times 10^{-14}~\msano$) by factors of $10$ to $10^4$ \citep{1992ApJ...397..225M, 1992SvA....36...70B, 1996ApJ...462L..91L, 1997A&A...319..578V, 2001ApJ...546L..57W}. Based on radio observations from IRAS, VLA and JCMT data, \citet{1992ApJ...397..225M} estimated mass-loss rates for active dM stars (in particular, YZ~CMi, Gl~644, Gl~873) of a few times $10^{-10}~\msano$. This result was contradicted by \citet{1996ApJ...462L..91L}, who claimed that the active M dwarfs YZ~CMi (dM4.5e) and AD~Leo (dM3.5e) should present $\dot{M} \lesssim 10^{-13}~\msano$, and later on by \citet{1997A&A...319..578V}, who suggested that the mass-loss rates of dM stars should be at maximum a couple of $\dot{M} \lesssim 10^{-12}~\msano$. \citet{1997A&A...319..578V} pointed out that both YZ~CMi and AD~Leo are considered young stars (ages of about $0.5$ -- $1$~Gyr), and that at their early life, $\dot{M}$ should have been about one order of magnitude larger: $\dot{M} \sim (1.1$ -- $3.6) \times 10^{-11}~\msano$. Even for the closest star to the Solar System, Proxima Centauri (dM5.5e), estimates of mass-loss rates are rather controversial. \citet{2002ApJ...578..503W} suggest an upper limit of $\dot{M} \lesssim 3 \times 10^{-13}~\msano$ and \citet{1996ApJ...460..976L} suggest $\dot{M} \lesssim 7 \times 10^{-12}(u_\infty/300~{\rm km~s}^{-1})~\msano$ (where $u_\infty$ is the wind terminal velocity), although \citet{2001ApJ...547L..49W} claim an upper limit of $\dot{M} \lesssim 4 \times 10^{-15}~\msano$, $5$ times below the value of the solar wind mass-loss rate. According to \citet{2010Icar..210..539Z}, if Proxima Cen is about $1$ -- $4$~Gyr old, the ``young'' Proxima Cen should have had considerably larger mass-loss rates $\dot{M} \sim (4$ -- $64) \times 10^{-11}~\msano$.
%\citet{2002ApJ...578..503W} also propose that, if Proxima Cen 

In this work, we investigate the coronal wind of a specific fully-convective dM star, \peg, for which observed surface magnetic maps have been acquired \citep{2006Sci...311..633D, 2008MNRAS.384...77M}. For this, we use three-dimensional (3D) magnetohydrodynamics (MHD) simulations based on our previous models developed for solar-like stars \citep{2009ApJ...699..441V} and weak-lined T Tauri stars \citep{2009ApJ...703.1734V,vidotto10}. For the first time, an observationally derived surface magnetic field map is implemented in MHD models of stellar winds for a low-mass star. \peg\ is a suitable case for modelling as a first step, because its surface magnetic distribution is close to potential, which implies that the adopted boundary conditions match the observed map closely. 
We self-consistently take into consideration the interaction of the outflowing wind with the magnetic field and vice-versa. Hence, from the interplay between magnetic forces and wind forces, we are able to determine the configuration of the magnetic field and the structure of the coronal winds.

This paper is organised as follows. In Section \ref{sec.model}, we present the details of the stellar wind model we use to describe the wind of \peg. Section \ref{sec.results} presents the results of our simulations. In Section \ref{sec.discussion}, we present the discussions of our work, including implications of the wind on habitability of planets. Section \ref{sec.conclusions} presents the conclusion of our work.

%%%%%%%%%%%%%%%%%%%%%%%%%%%%%%%%%%%%%%%%%%%%%%%%%%%%%%%%%%%%%%%%%%%%%%%%
\section{THE NUMERICAL STELLAR WIND MODEL}\label{sec.model}
Observations of the solar corona during eclipses have revealed the presence of closed magnetic loops, extending out to several solar radii. Beside the closed loops, regions of open magnetic field lines form coronal holes, from where the fast solar wind originates. Based on the observed solar configuration, bi-component coronal models, where closed and open magnetic field line regions coexist, have been used to explain the coronal structure of the Sun and other stars \citep{1971SoPh...18..258P, 1987MNRAS.226...57M, 1999A&A...343..251K, 2000ApJ...530.1036K, 2002ApJ...576..413U, 2008ApJ...678.1109M, 2009ApJ...699..441V, 2009ApJ...703.1734V, vidotto10}. In these models, closed field structures form a belt around the equator of the star, while open field structures give rise to the fast solar/stellar wind. We adopt such wind models to simulate the yet unobserved wind of \peg .

To perform the simulations, we use the 3D MHD numerical code BATS-R-US developed at University of Michigan \citep{1999JCoPh.154..284P}. BATS-R-US solves the MHD equations in the finite-volume scheme. It has currently several approximate Riemann solvers available, although originally used the upwind Roe-type scheme. The temporal integration is performed using either an explicit, implicit or combined explicit-implicit time stepping scheme \citep[details are found in][]{2006JCoPh.217..722T}. We perform our calculation in an explicit scheme, where a two-level Runge-Kutta scheme is used. Here, we summarise the characteristics of our model. More details can be found in \citet{vidotto10}. 

BATS-R-US solves the ideal MHD equations, that in the conservative form are given by
\begin{equation}
\label{eq:continuity_conserve}
\frac{\partial \rho}{\partial t} + \nabla\cdot \left(\rho {\bf u}\right) = 0,
\end{equation}
\begin{equation}
\label{eq:momentum_conserve}
\frac{\partial \left(\rho {\bf u}\right)}{\partial t} + \nabla\cdot\left[ \rho{\bf u\,u}+ \left(p + \frac{B^2}{8\pi}\right)I - \frac{{\bf B\,B}}{4\pi}\right] = \rho {\bf g},
\end{equation}
\begin{equation}
\label{eq:bfield_conserve}
\frac{\partial {\bf B}}{\partial t} + \nabla\cdot\left({\bf u\,B} - {\bf B\,u}\right) = 0,
\end{equation}
\begin{equation}
\label{eq:energy_conserve}
\frac{\partial\varepsilon}{\partial t} +  \nabla \cdot \left[ {\bf u} \left( \varepsilon + p + \frac{B^2}{8\pi} \right) - \frac{\left({\bf u}\cdot{\bf B}\right) {\bf B}}{4\pi}\right] = \rho {\bf g}\cdot {\bf u} ,
\end{equation}
where $\rho$ is the mass density, ${\bf u}$ the plasma velocity, ${\bf B}$ the magnetic field, $p$ the gas pressure, ${\bf g}$ the gravitational acceleration due to the central body, and $\varepsilon$ is the total energy density given by 
\begin{equation}\label{eq:energy_density}
\varepsilon=\frac{\rho u^2}{2}+\frac{p}{\gamma-1}+\frac{B^2}{8\pi} .
\end{equation}
We consider an ideal gas, so $p=n k_B T$, where  $k_B$ is the Boltzmann constant, $T$ is the temperature, $n=\rho/(\mu m_p)$ is the wind density, $\mu m_p$ is the mean mass of the particle, and $\gamma$ is the heating parameter (polytropic index). In practice, $\gamma$ represents the energy balance of the wind, for which the details are unknown. In our simulations, we adopt $\gamma=1.2$ and $\mu=0.5$.

BATS-R-US uses a Cartesian computational domain that is block-based, where blocks can be either created or destroyed as the computation proceeds. To do that, an adaptive mesh refinement (AMR) technique is used in the MHD algorithm of BATS-R-US. Although we opted not to modify our grid resolution during the numerical computation, we used the AMR capabilities of BATS-R-US to construct a grid for the simulations of the wind of \peg\ that is more refined around the star, which is placed at the centre of the Cartesian grid. The grid axes $x$, $y$, and $z$ extend from $-20~R_*$ to $20~R_*$, where $R_*$ is the stellar radius. After applying $8$ levels of refinement, the more refined zone of the grid (for $r<5~R_*$) has cells with a resolution of $7.8\times 10^{-2}~R_*$. 

The inner boundary of the system is the base of the wind at radius $r=R_*$, where fixed boundary conditions are adopted. The outer boundaries at the edges of the grid have outflow conditions, i.e., a zero gradient is set to all the primary variables (${\bf u}$, ${\bf B}$, $p$, and $\rho$).

\peg\ has mass $M_* = 0.28~M_\odot$, radius $R_*=0.34~R_\odot$ and is rotating with negligible differential rotation (i.e., as a solid body) with a period of rotation $P_0=0.44$~d \citep{2008MNRAS.384...77M}. We consider that its axis of rotation lies in the $z$-direction. The simulations are initialised with a 1D hydrodynamical wind for a fully ionised plasma of hydrogen. Immersed in this wind we consider a magnetic field anchored on the stellar surface that has either a dipolar geometry (\S\ref{dipolar}) or a geometry derived from extrapolations from observed surface magnetic maps (\S\ref{extrapolation}) using the potential-field source surface (PFSS) method. The MHD solution is evolved in time from the initial magnetic field configuration to a fully self-consistent solution. We do not adopt fixed topologies for either the magnetic field or for the wind, as both the wind and magnetic field lines are allowed to interact with each other.

%%%%%%%%%%%%%%%%%%%%%%%%%%%%%%%%%%%%%%%%%%%%%%%%%%%%%%%%%%%%%%%%%%%%%%%%
\subsection{Dipolar Surface Field}\label{dipolar}
\begin{figure*}
\includegraphics[width=84mm]{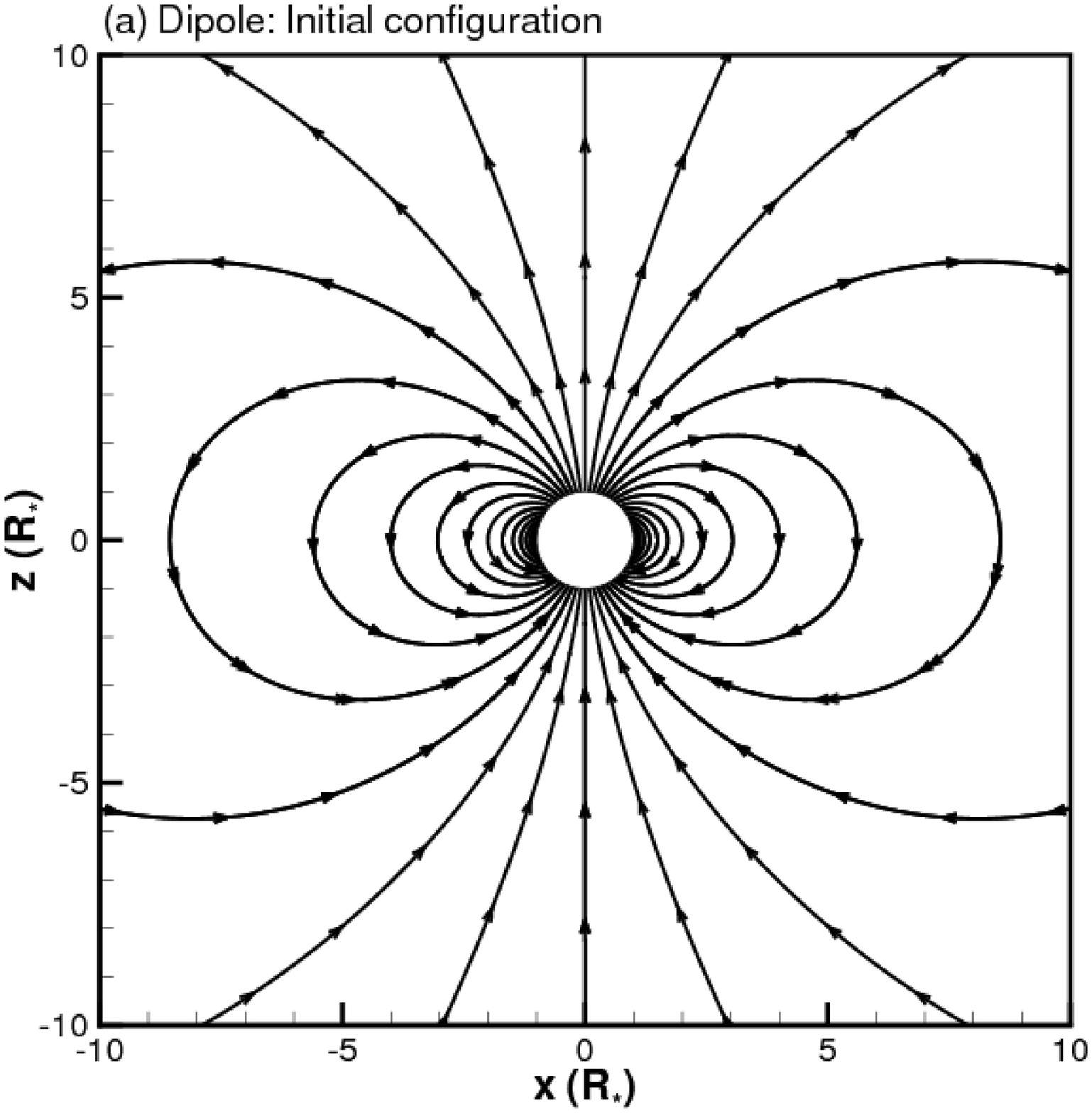}%
\includegraphics[width=84mm]{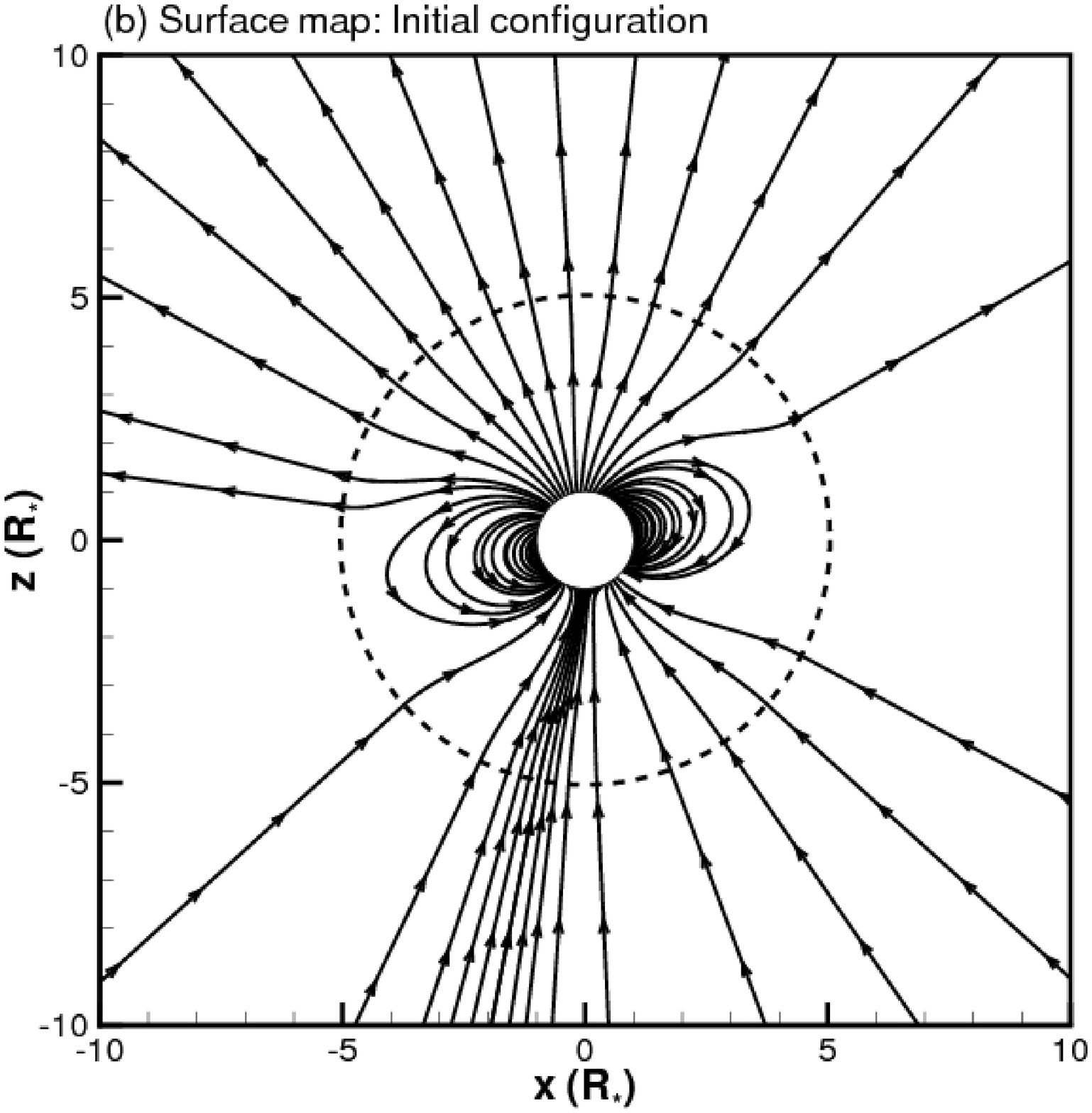}\\
\includegraphics[width=84mm]{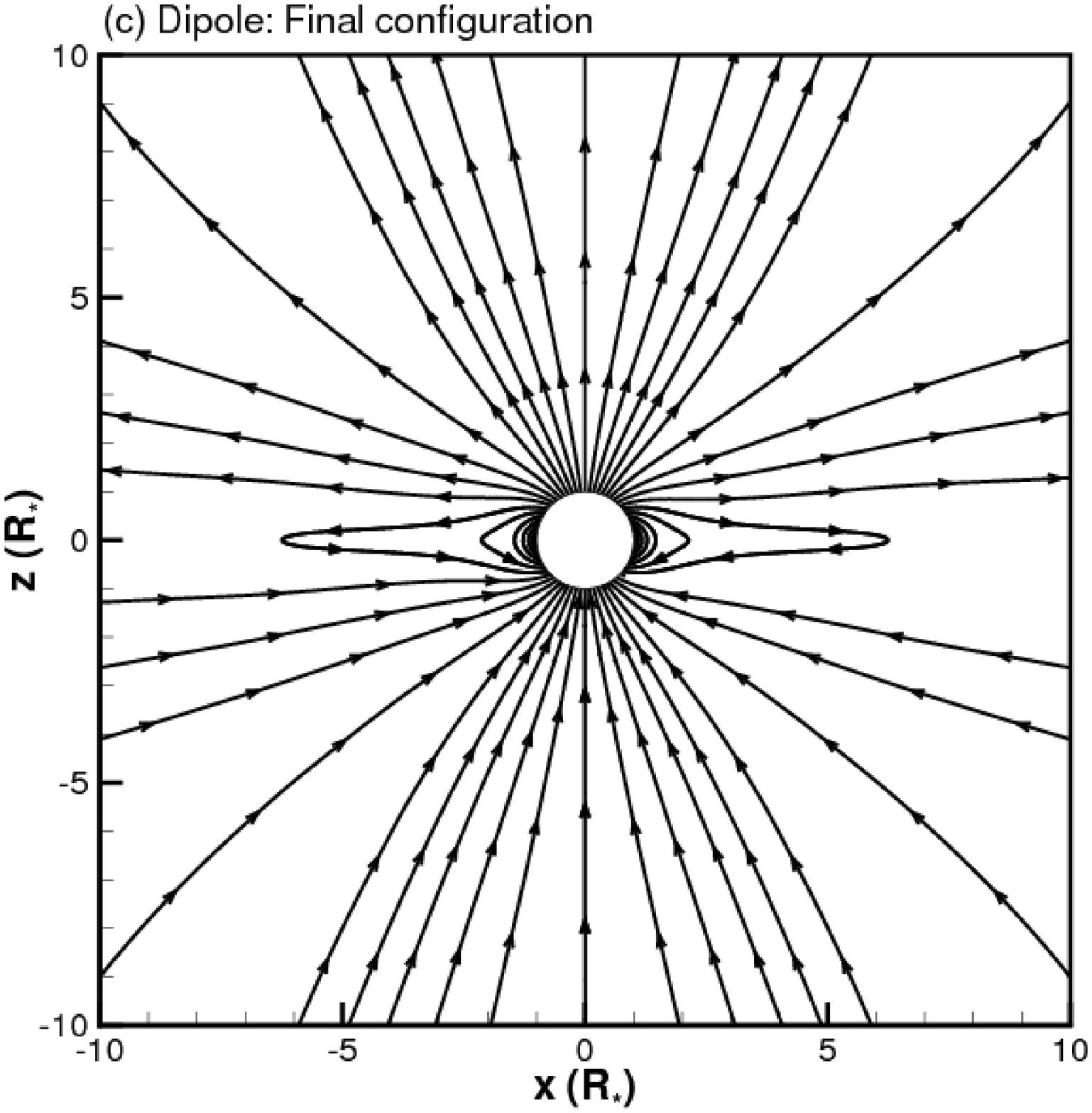}%
\includegraphics[width=84mm]{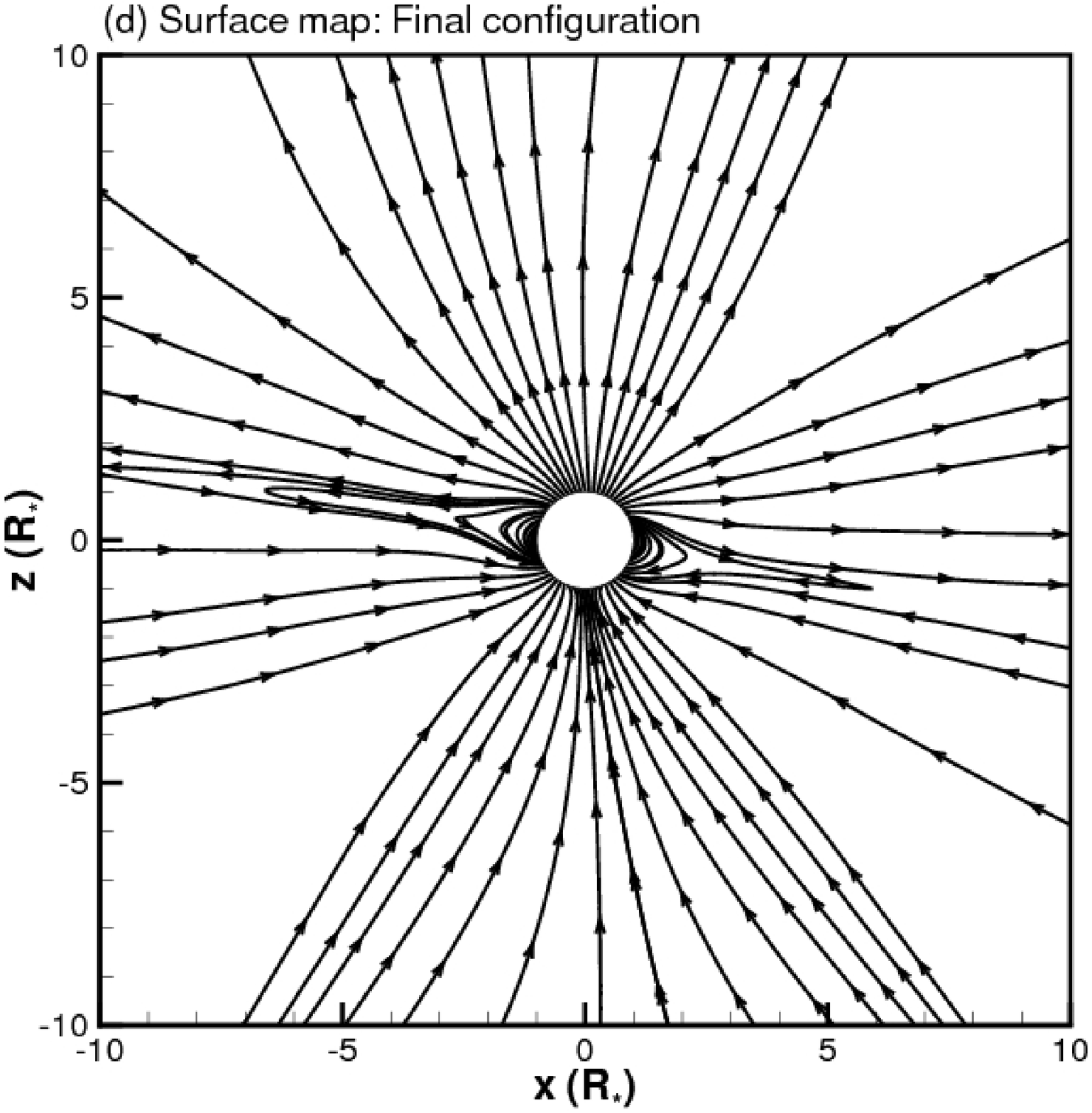}
\caption{(a) Initial configuration of magnetic field lines for a dipole and (b) for the extrapolation of the surface map using the potential-field source surface (PFSS) technique, where the dashed circumference represents the position of the source surface. (c) Final configuration of the of magnetic field lines after the self-consistent interaction with the stellar wind considering a dipolar surface distribution of magnetic fields and (d) considering the observationally derived magnetic map. Note that panels (b) and (d) are shown for different stellar rotation phases.  \label{fig.IC}}
\end{figure*}

Because \peg\ presents a surface magnetic field distribution that is mainly dipolar, a dipolar field geometry was initially used for our simulations. This allows comparisons with the previous works of \citet{2009ApJ...699..441V, 2009ApJ...703.1734V, vidotto10} and with a realistic magnetic field distribution at the stellar surface. In such cases, the initial magnetic field configuration is described in spherical coordinates $( r, \theta, \varphi )$ by
\begin{equation}
\label{eq:dipole}
{\bf B} =  \frac{B_0 R_* ^3}{r ^3} \left(\cos \theta , \frac12 \sin \theta, 0 \right).
\end{equation}
To be compatible to the maximum intensity derived in the surface maps, we adopted a maximum intensity of the magnetic field of $B_0=1660$~G, which is evaluated at the magnetic pole. Here, we assume that the dipole moment is aligned with the rotation axis of the star. Figure~\ref{fig.IC}a shows the initial configuration of the magnetic field lines.

%%%%%%%%%%%%%%%%%%%%%%%%%%%%%%%%%%%%%%%%%%%%%%%%%%%%%%%%%%%%%%%%%%%%%%%%
\subsection{Observed Surface Field}\label{extrapolation}
\begin{figure}
\includegraphics[width=84mm]{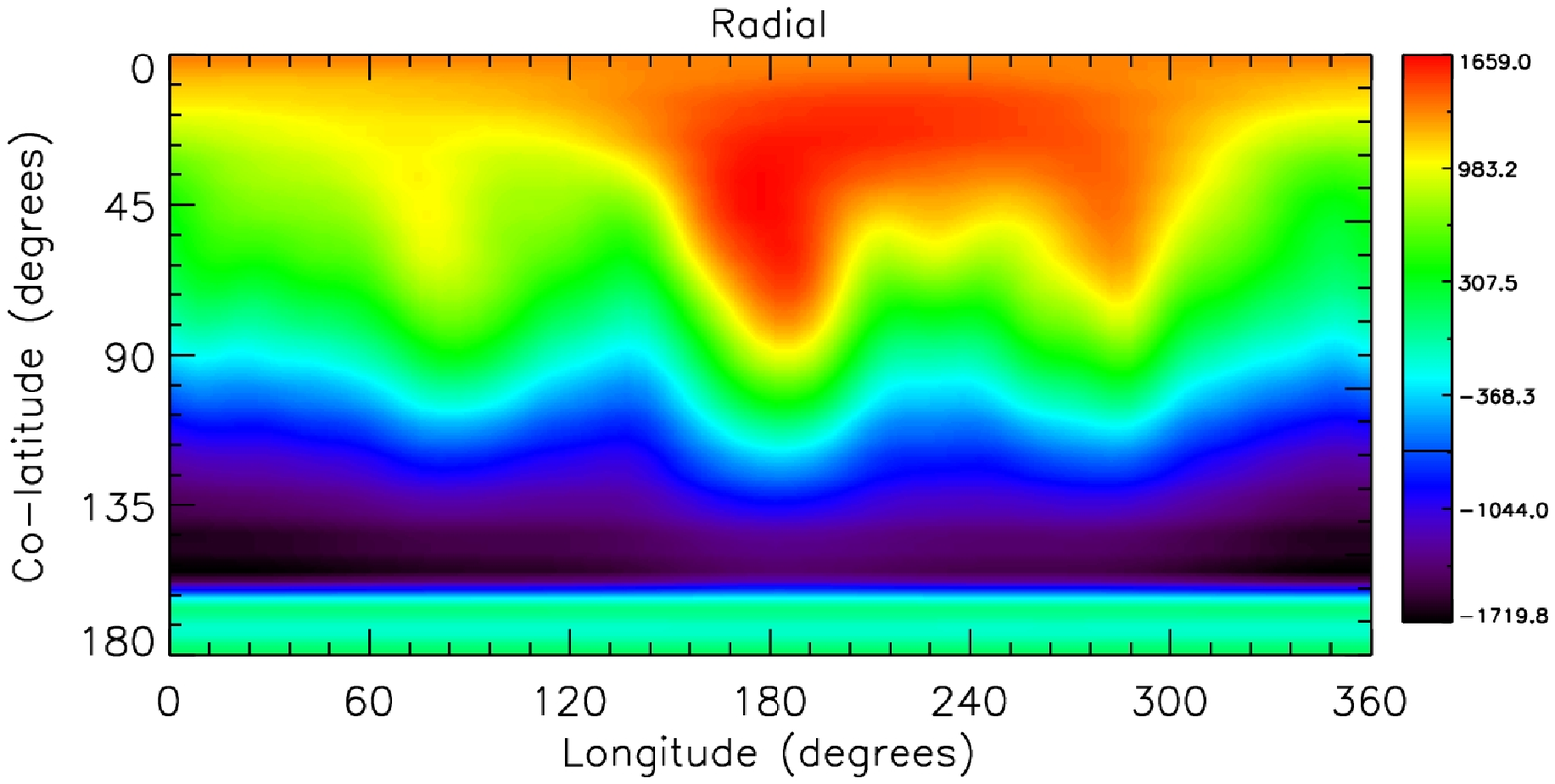}\\
\includegraphics[width=84mm]{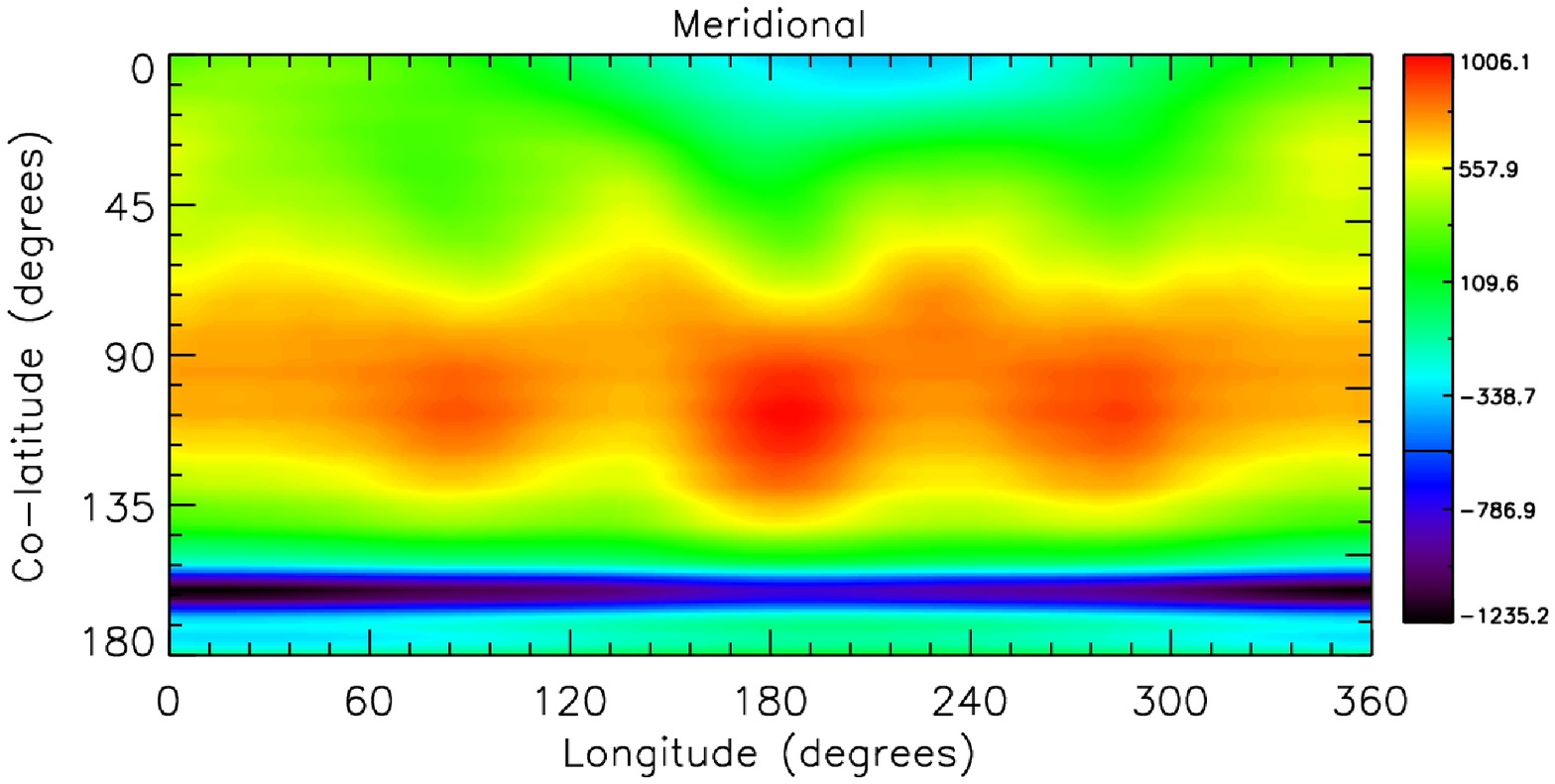}\\
\includegraphics[width=84mm]{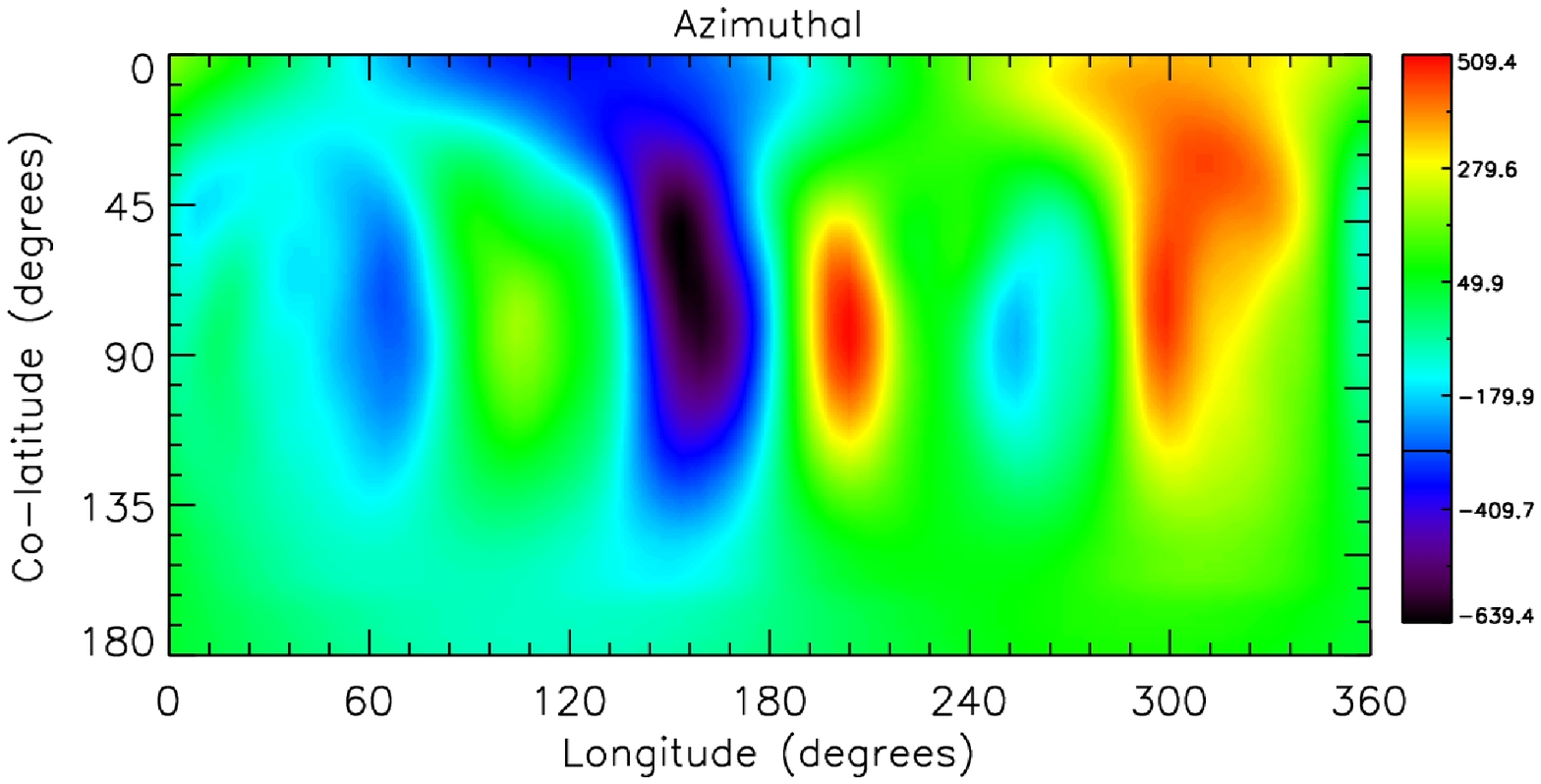}
\caption{The three components of the surface magnetic field of \peg\ obtained through extrapolations of the surface map using the PFSS technique. Maps were derived by \citet{2006Sci...311..633D, 2008MNRAS.384...77M} and extrapolations were done as described in \citet{2002MNRAS.333..339J}. \label{fig.maps}}
\end{figure}

In the remaining cases, we implemented a more realistic distribution of magnetic fields at the surface of the star. This is the first time an observed magnetogram has been included in a MHD wind model for a low-mass star. Magnetic maps for \peg\ have been presented in \citet{2006Sci...311..633D, 2008MNRAS.384...77M}, and are derived from data acquired in 2005 Aug, 2005 Sep, and 2006 Aug. As a first step, we have opted to implement the maps derived from the particular set of observations made in 2005 Aug. Figure~\ref{fig.maps} shows the radial, meridional and azimuthal components of the surface magnetic field of \peg\ obtained through extrapolations of the surface map using the PFSS technique, as described in \citet{2002MNRAS.333..339J}. In the PFSS model, the stellar wind plasma is not included directly, but its effects on the magnetic field (and vice-versa) are incorporated through the inclusion of the source surface. Such a surface, for instance, alters the number of open magnetic field lines, through where a stellar wind could escape. PFSS methods are usually criticised because their basic assumptions (the magnetic field is a potential field and the source surface is spherical) may not always be met. However, the advantage of the PFSS method over the MHD models relies on its simplicity: it is simpler to implement and requires much less computer resources than MHD models. In our model, we use the magnetic field configuration derived by the PFSS method as initial condition and boundary condition at the surface of the star. We note that the surface maps presented in Figure~\ref{fig.maps} matches the observed maps closely.

A meridional cut of the initial configuration of the magnetic field lines is presented in Figure~\ref{fig.IC}b. The source surface (dashed circumference) is chosen to lie at $r_{\rm SS} = 5~R_*$, where beyond that, the magnetic field is considered to be purely radial. We can see that the extrapolated field is indeed mainly dipolar, slightly inclined with respect to the rotational axis of the star ($z$-axis). We note that the surface of the star that occupies co-latitudes $\gtrsim 120^{\rm o}$ is never in view as the star rotates and so the magnetic field there can not be reconstructed reliably.

%%%%%%%%%%%%%%%%%%%%%%%%%%%%%%%%%%%%%%%%%%%%%%%%%%%%%%%%%%%%%%%%%%%%%%%%
\section{RESULTS}\label{sec.results}
Our simulations require a set of input parameters for the wind. Unfortunately, some of them are poorly constrained by observations. For \peg, the magnetic field is the better-constrained parameter. We have, therefore, implemented in our previous models \citep{vidotto10} surface magnetic maps derived by data acquired in 2005 Aug \citep{2006Sci...311..633D}. These observations show that \peg\ hosts an intense, mainly axi-symmetrical dipolar magnetic field, with maximum intensity of about $1660$~G, i.e., $3$ orders of magnitude larger than the Sun. 

The wind temperature and density are less constrained for \peg. We, therefore, adopt values representative of dM stars. dM stars are believed to host coronae with a high-temperature plasma $\sim 10^7$~K in conjunction with a low-temperature one $2$ -- $3 \times 10^6$~K \citep{1990ApJ...365..704S, 1996ApJ...463..707G}. In our models, we adopt a temperature at the base of the coronal wind of $T_0 = 2\times 10^6$~K or $10^7$~K. These coronal temperatures are about the same order of magnitude as the solar coronal temperatures of $1.56 \times 10^6~$K.

Coronal densities inferred from X-ray observations of dM stars suggest densities ranging from $10^{10}$~cm$^{-3}$ to $5 \times 10^{12}$~cm$^{-3}$ \citep{2002A&A...394..911N, 2004A&A...427..667N}. Therefore, we adopt, at the base of the coronal wind, densities in the range $10^{10}$ -- $10^{12}$~cm$^{-3}$. Compared to the solar coronal density of about $\sim 2 \times 10^8$~cm$^{-3}$, coronal densities inferred for dM stars are about $2$ -- $4$ orders of magnitude larger than for the solar corona.

The density, along with the magnetic field, are key parameters in defining the magnetic field configuration of the stellar wind and its velocity profile \citep{2009ApJ...699..441V,2009ApJ...703.1734V}. Together, they define the plasma-$\beta$, defined by the ratio of thermal to magnetic energy densities. Therefore, at the base of the coronal wind of \peg, 
\begin{equation}\label{eq.beta0}
\beta_0 = \frac{n_0 k_B T_0}{B_0^2/(8\pi)} \simeq 2.5 \times 10^{-5} n_{10} T_6, 
\end{equation}
where the index ``0'' means the variable is evaluated at the base of the coronal wind, $n_{10} = n_0/(10^{10}$~cm$^{-3})$ and $T_6=T_0/(2\times10^6$~K). For $n_{10}=1$ and $T_6=1$, $\beta_0$ is about $5$ orders of magnitude smaller than for the solar wind \citep{1971SoPh...18..258P}. This implies that the winds of dM stars are highly magnetised and, therefore, are expected to differ from solar-type winds. 

Table~\ref{table} presents the parameters adopted for the set of simulations we performed. The suffixes `Dip' and `Map' at the case name (first column) stand for the distribution of the surface magnetic field adopted: dipolar or from a magnetic map, respectively. The dipolar cases form the Set 1 of simulations while the cases with magnetograms form the Set 2.

\begin{table*} 
\centering
%\begin{minipage}{84mm}
\caption{Adopted parameters for the sets of simulations. The columns are, respectively: the case name, the density $n_0$ and temperature $T_0$ at the base of the coronal wind ($r=R_*$), the plasma-$\beta$ parameter evaluated at $R_*$ [Eq.~\ref{eq.beta0}], the mass loss rate $\dot{M}$, the angular momentum loss rate $\dot{J}$, and the time-scale for rotational braking $\tau$. The case names are labelled either ``Dip'' or ``Map'' to indicate that the distribution of the surface magnetic field is, respectively, dipolar or from an observed magnetogram.\label{table}}   
\begin{tabular}{c c c  c c c c }  
\hline\hline    
{Case} &  {$n_0$}  & {$T_0$} &    {$\beta_0$} &  {$\dot{M}$}  & {$\dot{J}$} &   {$\tau$} \\
  & (cm$^{-3}$) & (MK)      &  & ($10^{-11}$~M$_\odot~{\rm yr}^{-1}$) & ($10^{33}$~erg~s$^{-1}$) &  (Myr) \\
\hline
\multicolumn{7}{c}{Set 1:    Dipole }  \\
\hline
1Dip &    $10^{10}$ & $2$  &    $2.52\times10^{-5}$ &  $3.4$ & $2.3$ &  $280$ \\
2Dip &    $10^{11}$ & $2$  &    $2.52\times10^{-4}$ & $12$ & $7.7$ &   $84$  \\
3Dip &    $10^{12}$ & $2$  &    $2.52\times10^{-3}$ & $41$ & $26$ &   $24$   \\
4Dip &    $10^{11}$ & $10$ &    $1.26\times10^{-3}$ & $25$ & $8.3$ &   $78$  \\
\hline    
\multicolumn{7}{c}{Set 2:    Map } \\
\hline
1Map  &    $10^{10}$ & $2$  &    $2.52\times10^{-5}$ &    $4.2$ & $3.4$ &   $180$    \\
2Map  &    $10^{11}$ & $2$  &    $2.52\times10^{-4}$ &    $14$ & $7.6$ &   $84$   \\
3Map  &    $10^{12}$ & $2$  &    $2.52\times10^{-3}$ &    $50$ & $32$ &   $17$   \\
4Map  &    $10^{11}$ & $10$ &    $1.26\times10^{-3}$ &    $26$ & $9.1$ &   $48$    \\
\hline                              
\end{tabular}
%\end{minipage}
\end{table*}

We were able to find a MHD solution for the wind for all the simulations we ran, showing that it is possible to develop coronal wind models with a realistic distribution of magnetic field. In general, MHD wind models are studied under the assumption of simplistic magnetic field configurations, especially when in pursuit of an analytical solution. Therefore, the study of a magnetised coronal wind where an observed magnetic field distribution is considered has long been awaited. Our work also sheds some light on the yet unobserved winds from dM stars.

Our results show that cases where all parameters are the same, except for the surface magnetic field configuration (e.g., 1Dip/1Map, or 2Dip/2Map, and so on) present similar general wind characteristics and magnetic field configurations. The wind velocities are approximately the same for cases 3Dip and 3Map, although the details of the velocity profiles may differ. Overall, our solutions differ considerably from the solar wind solution, where a low-velocity wind (terminal velocities of $u_{\infty , \odot} \simeq 400$ -- $800$~km~s$^{-1}$) with low mass-loss rate ($\dot{M}_\odot \simeq 2 \times 10^{-14}~\msano$) is found. Terminal velocities of our models are around $u_\infty \simeq 1500$ -- $2300$~km~s$^{-1}$ for cases where $n_0 = 10^{12}$~cm$^{-3}$, increasing as $n_0$ decreases, while the mass-loss rates are about $\dot{M} \approx 4 \times 10^{-10}~\msano$ increasing as $n_0$ increases. A comparison of terminal velocities derived from simpler wind models is presented in Appendix~\ref{ap.windmodels}. We note that, based on more simplistic wind models, such as \citet{1967ApJ...148..217W}, in the fast magnetic rotator limit, wind terminal velocities of $\simeq 3320~{\rm km~s}^{-1}$ are expected for a wind mass-loss rate of about $10^{-11}~\msano$. Figures~\ref{fig.IC}c and \ref{fig.IC}d present the final configuration of the magnetic field lines, where we note the similarities of the field topology for cases with a dipolar distribution of surface magnetic flux (Fig.~\ref{fig.IC}c) and with a distribution given by observed magnetic maps (Fig.~\ref{fig.IC}d). As a result of the rotation of the star, the apparent positions of the magnetic poles in the initial (Fig.~\ref{fig.IC}b) and in the final magnetic field configurations (Fig.~\ref{fig.IC}d) are different because the  plots represent two different stellar phases of rotation.

\citet{2010MNRAS.tmp.1077M} suggested that the magnetic field structure of very low-mass stars may switch from a strong, simple field configuration to a weak, complex field over time. If this is the case for \peg\ and its magnetic field is in such a transient state, its wind characteristics as well as mass-loss and angular momentum-loss rates are expected to change along with its magnetic field structure.

In the next sections, we present more details of our results divided in two sets defined by the configuration of the surface magnetic distribution adopted.

%%%%%%%%%%%%%%%%%%%%%%%%%%%%%%%%%%%%%%%%%%%%%%%%%%%%%%%%%%%%%%%%%%%%%%%%
\subsection{Set 1: Dipolar Field}\label{sub.resdipolar}
Set 1 considers a magnetic surface distribution of a dipolar field. By comparing cases 1Dip, 2Dip, and 3Dip, i.e., where only the base coronal density $n_0$ was varied, we found that the poloidal velocity of the wind scales approximately as
\begin{equation}\label{eq.windlaw}
u_p^2 \propto \frac{1}{n_0} {\rm ~for~a~given~}B_0.
\end{equation}
This qualitatively agrees with previous results \citep{2009ApJ...699..441V}, where it was found that an increase in the density leads to winds with lower velocities. Simulations 1Dip, 2Dip, and 3Dip are in a very low-$\beta$ regime, where magnetic effects completely override thermal and kinematic effects of the wind. The equation of motion in steady state reduces to
\begin{equation}\label{eq.navier-stokes_eulerian}
\rho ({\bf u} \cdot \nabla ) {\bf u} = - {\bf \nabla} p + \rho {\bf g} +   \frac{{\bf j} \times{\bf B}}{c} ,
\end{equation}
where ${\bf j} ={c}({\bf \nabla} \times {\bf B})/({4 \pi})$ is the current density. For simplicity, let us take the radial component of previous equation and integrate it. Neglecting terms with $p$ and $\rho{\bf g}$ relative to the much larger magnetic term, the previous equation is rewritten as
\begin{equation}
u_r^2 \sim \int \frac{|{\bf j} \times{\bf B}|_r}{c \rho} {\rm d}r .
\end{equation}
This shows that in a domain where $p\ll B^2$ (very low-$\beta$ regime) and gravitational forces can be neglected, the squared velocity of the wind is inversely proportional to the density [Eq.~(\ref{eq.windlaw})].

\begin{figure*}
\includegraphics[width=84mm]{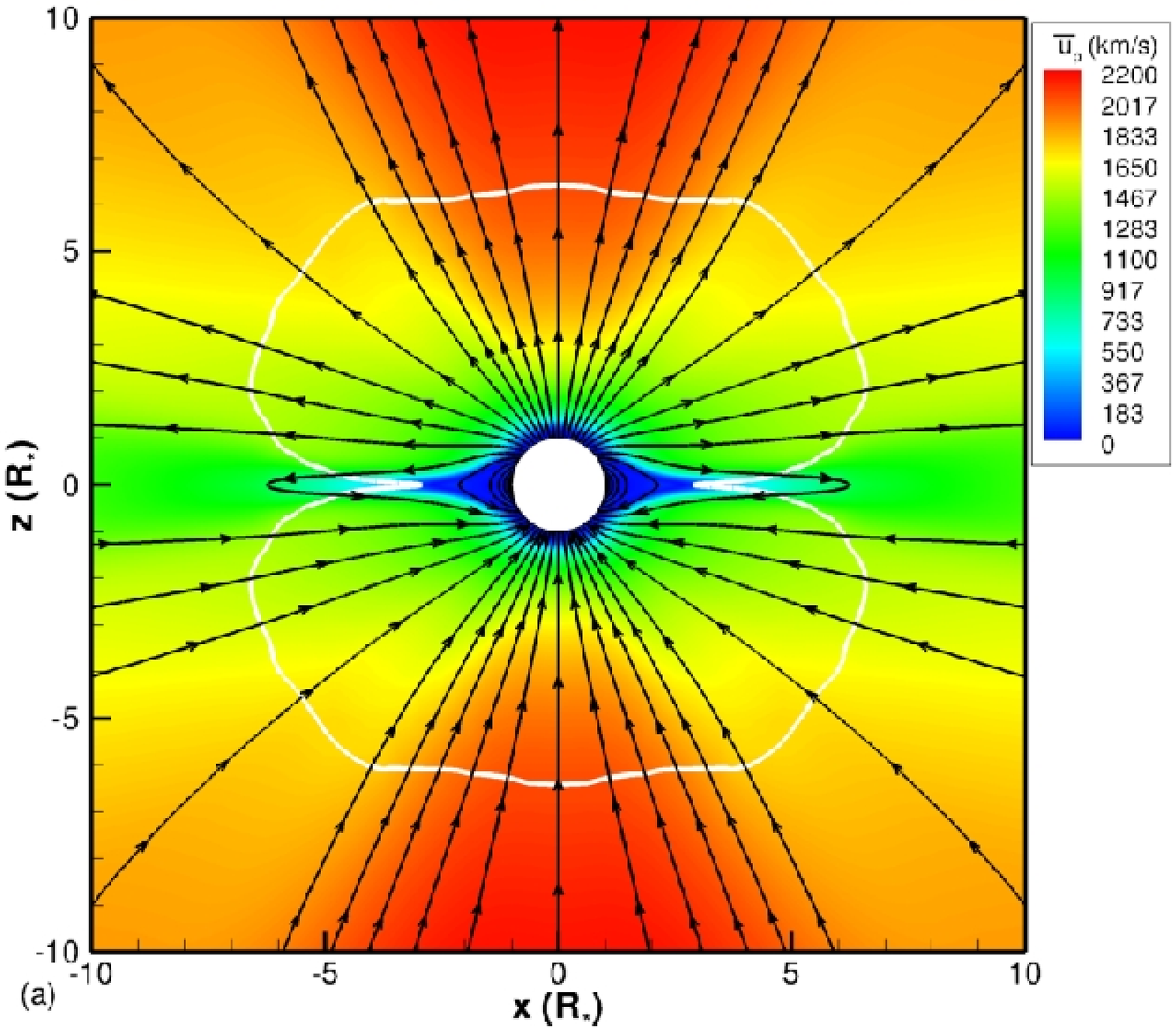}%
\includegraphics[width=84mm]{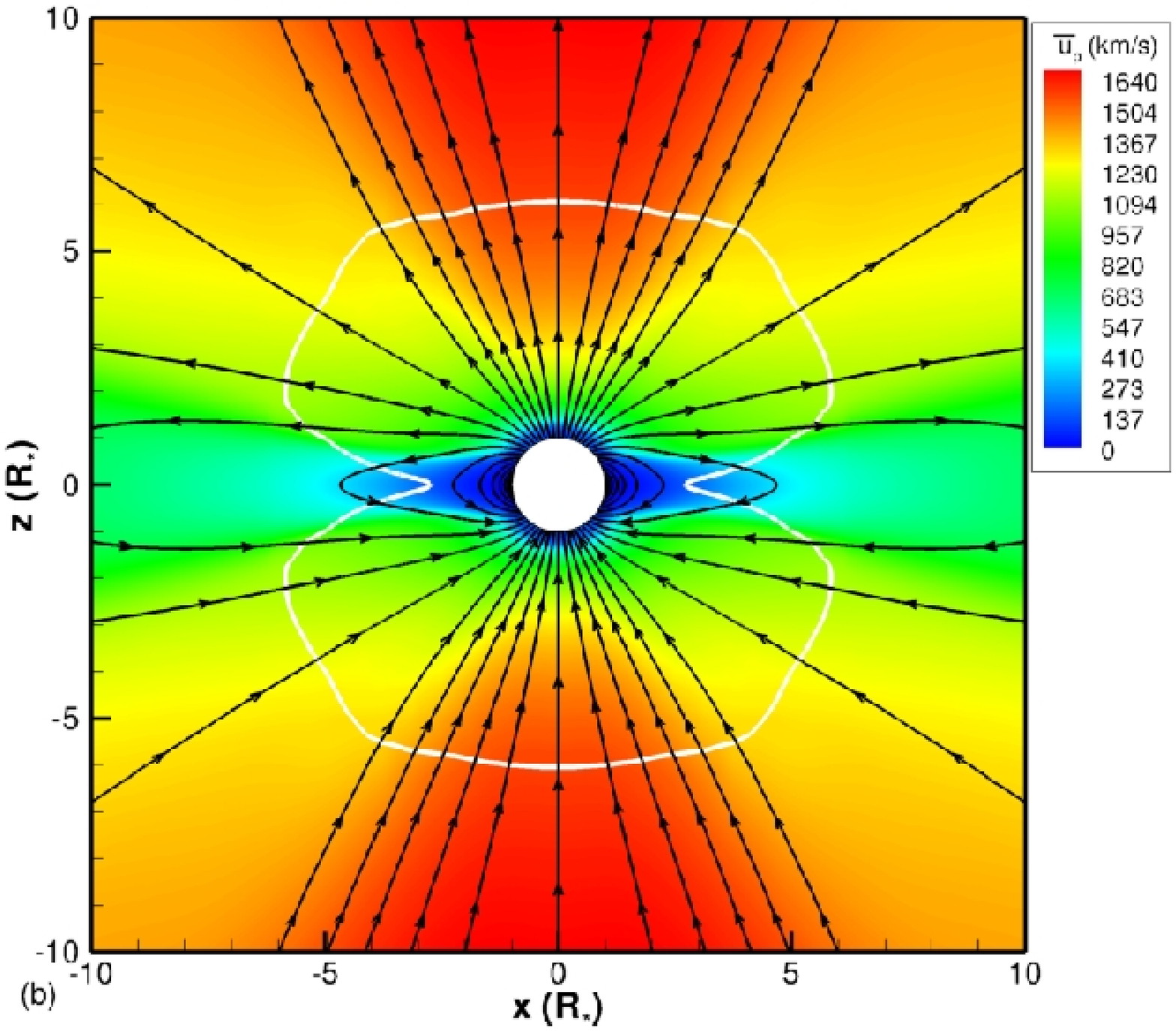}\\
\includegraphics[width=84mm]{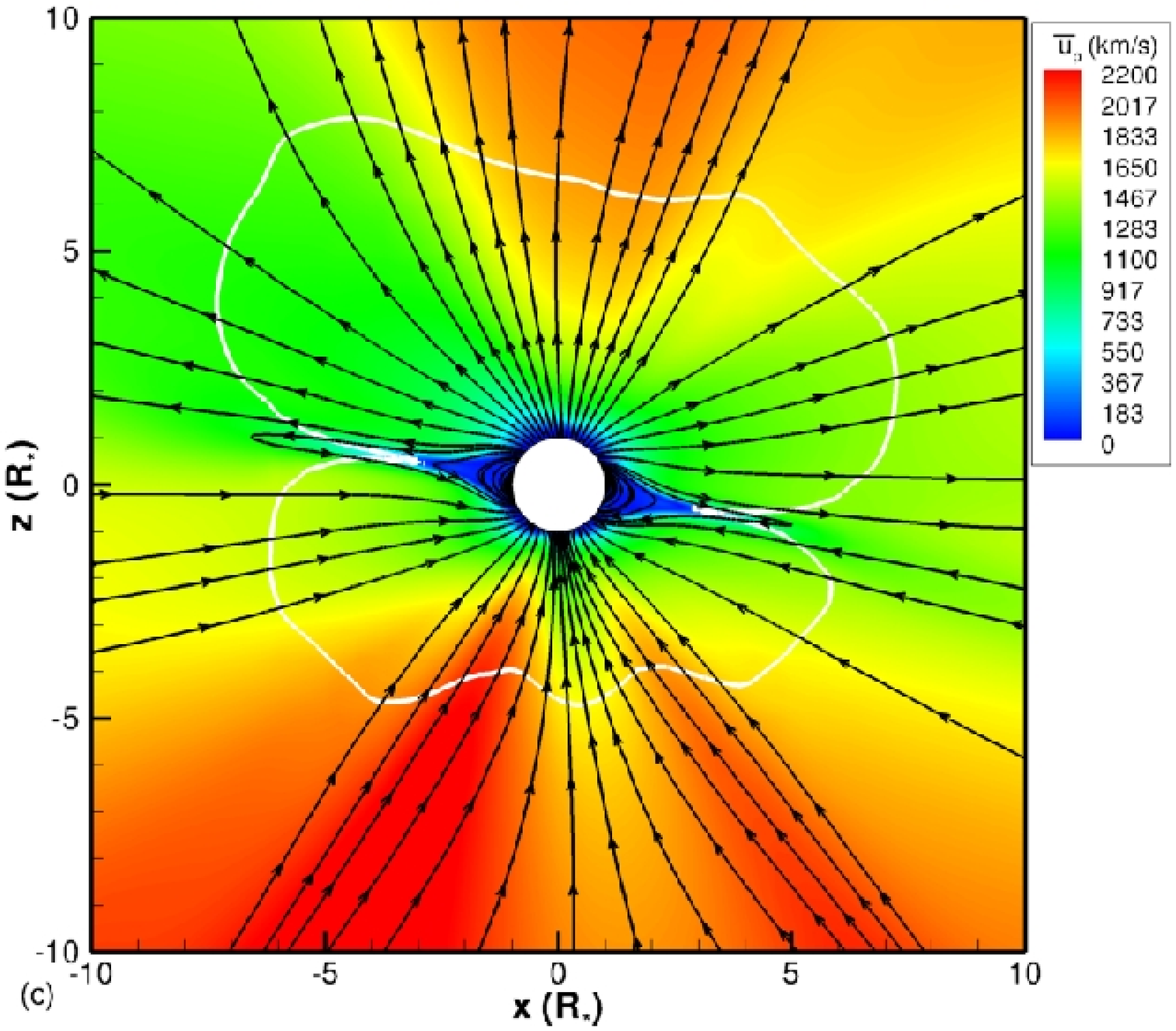}%
\includegraphics[width=84mm]{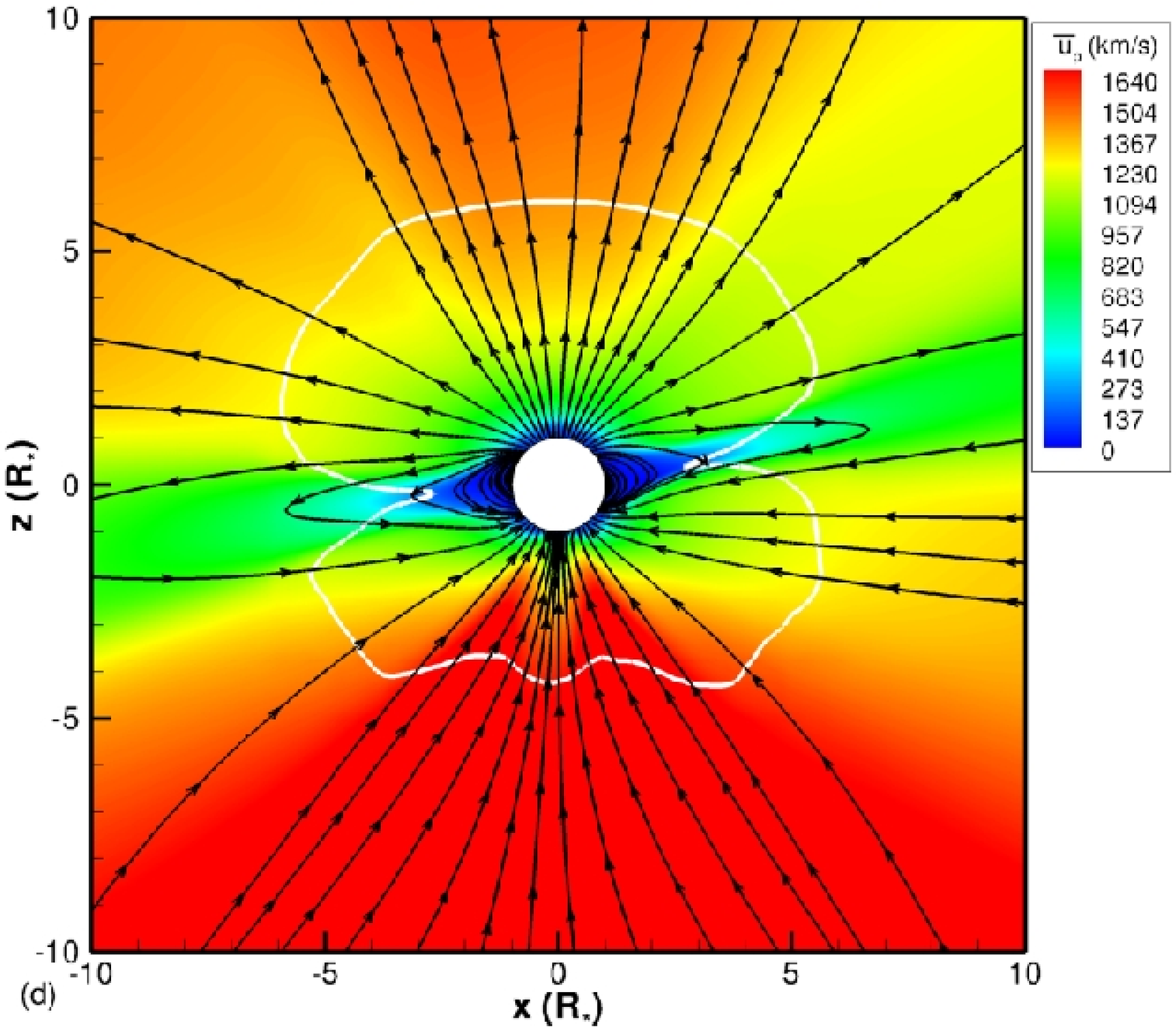}
 \caption{Meridional cut of scaled poloidal wind velocity $\bar{u}_p=u_p/\sqrt{n_{12}}$ profile (a) for cases with a dipolar field surface distribution 1Dip, 2Dip, and 3Dip, and (b) 4Dip, (c) for cases where a magnetic surface map was used 1Map, 2Map, and 3Map, and (d) 4Map. Black lines represent the magnetic field configuration, and white line is the location of the \alf\ surface. \label{fig.windvel}}
\end{figure*}

At the \alf\ surface, the poloidal wind velocity equals the local \alf\ velocity
\begin{equation}
u_{p,A}^2 = \frac{B_{p,A}^2}{4\pi\rho_A} .
\end{equation}
As $u_{p,A}^2 \propto 1/\rho_A$ for cases 1Dip, 2Dip, and 3Dip [Eq.~(\ref{eq.windlaw})], the location of the \alf\ surface will only depend on the intensity $B_{p,A}$ of the magnetic field at the \alf\ surface. As these cases possess the same magnetic field at the base of the coronal wind, and because the configuration of the magnetic field lines is very similar, the magnetic field at the \alf\ surface should be approximately the same. This implies that, despite the increase in the base density, the location of the \alf\ surface is similar for cases 1Dip, 2Dip, and 3Dip. This is different to what was found in \citet{2009ApJ...699..441V}, where an increase in the coronal density modified the location of the \alf\ surface. This {\it \alf\ surface saturation} happens because simulations 1Dip, 2Dip, and 3Dip are in a very low-$\beta$ regime. Therefore, Eq.~(\ref{eq.windlaw}) should be treated with caution, as under different $\beta$ regimes (for example, when it approaches $\beta_0\sim 1$), it becomes invalid. Figure~\ref{fig.windvel}a presents the scaled wind velocity profile $\bar{u}_p$ for cases 1Dip, 2Dip, and 3Dip,
\begin{equation}\label{eq.scaledv}
\bar{u}_p = {u_p}{n_{12}^{1/2}} ,
\end{equation}
where $n_{12} = n_0 /(10^{12}$~cm$^{-3}$). We note that the wind terminal velocity is $u_\infty \approx (1500$ -- $2300) n_{12}^{-1/2}~{\rm km~s}^{-1}$, where the range of velocities refers to different wind latitudes (low-wind velocity near the equator, high-wind velocity around the poles). 

Equation~(\ref{eq.windlaw}) also implies that the mass-loss rate of the wind ($\dot{M} \propto \rho u_r $) should scale as
\begin{equation}\label{eq.windmassloss}
\dot{M} \propto \rho u_r \propto  n_0^{1/2},
\end{equation}
which means that, despite the fact that the wind velocity of case 3Dip is $10$ times smaller than case 1Dip [Eq.~(\ref{eq.windlaw})], its mass-loss rate is one order of magnitude larger than for case 1Dip [Eq.~(\ref{eq.windmassloss})]. This has implications on the efficiency of angular momentum loss, as will be shown in \S\ref{subsec.angmom}. The mass-loss rates for cases 1Dip, 2Dip, and 3Dip are $\dot{M} \approx 4\times 10^{-10} n_{12}^{1/2}~\msano$. 

Figure~\ref{fig.windvel}b shows the scaled poloidal velocity profile $\bar{u}_p$ for case 4Dip. This case considers a different temperature at the base of the corona ($10^7$~K as opposed to $2\times10^6$~K), and, because of that, has a larger $\beta_0$ (Table~\ref{table}). For this case, we did not find an analytical expression relating velocity and temperature. The \alf\ surface location and configuration of magnetic field lines are similar to the other dipolar cases, but the scaled wind velocity $\bar{u}_p$ is smaller than the previous cases: $u_\infty \approx (800$ -- $1800) n_{12}^{-1/2}~{\rm km~s}^{-1}$. The lower velocity observed in case 4Dip happens because of its higher $\beta_0$. The mass-loss rate for case 4Dip is $\dot{M} \simeq 2.5\times 10^{-10}~\msano$.

%%%%%%%%%%%%%%%%%%%%%%%%%%%%%%%%%%%%%%%%%%%%%%%%%%%%%%%%%%%%%%%%%%%%%%%%
\subsection{Set 2: Observationally Derived Magnetic Map}\label{sub.magnetogram}
The results of the simulations where the magnetic field distribution at the surface of the star is set according to observed magnetic maps is presented in a similar fashion as done in \S\ref{sub.resdipolar}: the profile of the scaled velocity $\bar{u}_p$ [Eq.~(\ref{eq.scaledv})] for cases 1Map, 2Map, and 3Map is shown in Figure~\ref{fig.windvel}c, while for case 4Map it is presented in Figure~\ref{fig.windvel}d. Because the magnetic field in the lower hemisphere of the star is not reliably reconstructed (co-latitudes $\gtrsim 120^{\rm o}$ of the surface of the star are not observed), a high-velocity wind develops there. Although this feature is local and does not affect the remaining parts of the grid other than radially away from the stellar surface, it is an artifact of our method and should not be taken into consideration (e.g., see the accumulation of magnetic field lines near the low-hemisphere of the star in Figure~\ref{fig.IC}b). The wind velocities of Set 2 are consistent with the wind velocities of Set 1, shown in \S\ref{sub.resdipolar}. This can be seen by comparing Fig.~\ref{fig.windvel}a and \ref{fig.windvel}c, and \ref{fig.windvel}b and \ref{fig.windvel}d. We found similar values of wind terminal velocities ($u_\infty \approx [1300$ -- $2100] n_{12}^{-1/2}~{\rm km~s}^{-1}$ for cases 1Map, 2Map, and 3Map, and $u_\infty \approx [850$ -- $1600] n_{12}^{-1/2}~{\rm km~s}^{-1}$ for case 4Map) and mass-loss rates ($\dot{M} \approx 4\times 10^{-10} n_{12}^{1/2}~\msano$ for cases 1Map, 2Map, and 3Map, and $\dot{M} \simeq 2.6\times 10^{-10}~\msano$ for case 4Map). This similarity occurs because the observed magnetic map does not deviate significantly from the dipolar field adopted in \S\ref{sub.resdipolar}. Future works will consider other stars with more complex surface magnetic field distributions, where more significant deviations are expected.

Tests with different positions for the source surface showed that final configuration of the wind does not differ from what is presented in this paper.

%%%%%%%%%%%%%%%%%%%%%%%%%%%%%%%%%%%%%%%%%%%%%%%%%%%%%%%%%%%%%%%%%%%%%%%%
\section{DISCUSSION}\label{sec.discussion}
%%%%%%%%%%%%%%%%%%%%%%%%%%%%%%%%%%%%%%%%%%%%%%%%%%%%%%%%%%%%%%%%%%%%%%%%
\subsection{Constraining Model Parameters: Angular Momentum Loss}\label{subsec.angmom}
One way to restrict the parameters of our model is by comparing the derived values for $u_\infty$ or $\dot{M}$ with observed values. However, traditional measurements of mass-loss rates or wind terminal velocities have not yet been precisely obtained for dM stars. Estimates of mass-loss rates from dM stars, for instance, are rather controversial, ranging from $10$ to $10^4$ times the solar wind mass-loss rate \citep[e.g.,][]{1992ApJ...397..225M, 1992SvA....36...70B, 1996ApJ...462L..91L, 1997A&A...319..578V, 2001ApJ...546L..57W}, but predictions for sub-solar values exist as well \citep{2001ApJ...547L..49W}. Mass-loss rates derived from our wind models are $\dot{M} \approx 4\times 10^{-10} n_{12}^{1/2}~\msano$ for cases 1Dip/1Map, 2Dip/2Map, and 3Dip/3Map, and $\dot{M} \simeq 2.5\times 10^{-10}~\msano$ for case 4Dip/4Map. Such values are in accordance with several estimates. Unfortunately, they do not allow us to constrain the input parameters of our model, in particular, the density at the coronal base.

Another possible way to restrict our model parameters is to compare the time-scale for rotational braking $\tau$ with the age $t$ of \peg. Because \peg\ is a fast rotator, we expect that $\tau \gg t$. However, there exist only loose age estimates for \peg. \citet{1998A&A...331..581D} and \citet{2003ApJ...583..451M} estimated the kinematic age of \peg\ as ``young disc'', which roughly corresponds to $t \lesssim 3$~Gyr. Using the calibration of $L_X$ versus age provided by \citet{2009IAUS..258..307C}, for the observed value of $L_X = 10^{28.44}$~erg~s$^{-1}$ \citep{1998A&A...331..581D}, it results in an age estimate of $t \sim 0.7$ -- $0.8$~Gyr. The calibration from \citet{2009IAUS..258..327W} for M4 dwarfs results in the considerably older age of $t\sim 5.4$~Gyr, using $L_{{\rm H}_{\alpha}}$ and $L_{\rm bol}$ as given by \citet{1998A&A...331..581D}. Because age estimates are still loose and imprecise, its comparison with $\tau$ is not able to restrict our model parameters. 

Observations of the rotation evolution of dM stars in open clusters at different ages, on the other hand, provide a way to constrain the time-scale $\tau$ for the angular-momentum loss. It has been suggested that $\tau \sim 200$~Myr or, mostly likely $400$ -- $800$~Myr, \citep{2007MNRAS.381.1638S} for dM stars.

Angular momentum of the star is carried away by the stellar wind. Because in some of our simulations there is no axi-symmetry, the torque $\dot{\bf J}$ on the star has $x$, $y$ and $z$ components. Here, we are interested only on the $z$-component, as it is the only one responsible for the rotational braking (because the angular velocity of the star points in the $z$-direction). The $z$-component of the angular momentum carried by the wind is \citep{1970MNRAS.149..197M}
\begin{eqnarray}\label{eq.jdot}
\dot{J} &=&  \left[\alpha{\bf \hat{z}} \times \int_{V_A}  {\bf r} \times \rho ({\bf V }+ \alpha{\bf \hat{z}} \times {\bf r}){\rm d}V_A \right]_z \nonumber \\&+& \int_{S_A} \left( p + \frac{B^2}{8\pi} \right) ({\bf r} \times {\bf \hat{n}})_z {\rm d}S_A \nonumber \\ &+& \int_{S_A} \left[ {\bf r} \times (\alpha{\bf \hat{z}} \times {\bf r})\right]_z \rho {\bf V} \cdot {\bf \hat{n}} {\rm d} { S_A} ,
\end{eqnarray} 
where ${\bf V} = {\bf u} - \alpha {\bf \hat{z}}\times {\bf r}$ is the velocity vector in the frame rotating with angular velocity $\alpha {\bf \hat{z}}$, ${\bf \hat{z}}$ is the unit vector that points in the $z$-direction, $S_A$ is the \alf\ surface that delimits the volume $V_A$, and ${\bf \hat{n}}$ is the normal unit vector to the \alf\ surface\footnote{The computation of the local unit vector ${\bf \hat{n}}$ normal to the \alf\ surface was done with the routine developed by F.~Alouani~Bibi \citep{2009Natur.462.1036O}.}. The first term on the right of Eq.~(\ref{eq.jdot}) does not contribute to the $z$-component torque and is therefore null. The second term disappears in the case of a spherical \alf\ surface. It is also null in the axi-symmetric cases considered here (i.e., the dipolar cases), but it is non-null in the cases where a surface magnetic map is considered and it becomes relatively more important for the cases with larger adopted $\beta_0$. The third term is the dominant term in our simulations. 

We can estimate the time-scale for rotational braking as $\tau = {J}/{\dot{J}}$, where $J$ is the angular momentum of the star. If we assume a spherical star with a uniform density, then $J = 2/5 M_* R_*^2 \Omega_0$ and the time-scale is
\begin{equation}
\tau \simeq \frac{9 \times 10^{36}}{\dot{J}} \left( \frac{M_*}{M_\odot}\right) \left( \frac{1~{\rm d}}{P_0} \right) \left(  \frac{R_*}{R_\odot}\right)^2 ~{\rm Myr},
\end{equation} 
where $P_0=2\pi/\Omega_0$ is the rotational period of the star. For \peg , this results in 
\begin{equation}\label{eq.tau}
\tau \simeq \frac{6.45 \times 10^{35}}{\dot{J}} ~{\rm Myr}.
\end{equation} 

Because $\dot{J}$ depends on the mass flux crossing a given surface, i.e., on the mass-loss rate of the wind $\dot{M}$, from Eq.~(\ref{eq.windmassloss}), we have a rough scaling relation between $\dot{J}$ and $\dot{M}$ for cases 1Dip/1Map, 2Dip/2Map, and 3Dip/3Map
\begin{equation}\label{windangloss}
\dot{J} \propto \dot{M} \propto n_0^{1/2}  ,
\end{equation} 
which implies in a time-scale [Eq.~(\ref{eq.tau})] for rotational braking that scales as 
\begin{equation}\label{eq.windtimescale}
\tau \propto n_0^{-1/2} .
\end{equation} 
For cases 1Dip, 2Dip, and 3Dip, this implies that $\tau \simeq 28 {n_{12}^{-1/2}}$~Myr, while for cases 1Map, 2Map, and 3Map, $\tau \simeq 18 {n_{12}^{-1/2}}$~Myr. In either case, $\tau$ is well below the estimated solar spin-down time $\tau_\odot \simeq 7$~Gyr \citep{1967ApJ...148..217W}.

Table~\ref{table} presents the mass and angular momentum loss rates, and the time-scale for rotational braking calculated for all simulations, where we verify the approximate scaling given by Eqs.~(\ref{eq.windmassloss}), (\ref{windangloss}), and (\ref{eq.windtimescale}). Comparing to the observationally derived rotational braking time-scales of a couple of hundreds of Myr for dM stars is open clusters \citep{2007MNRAS.381.1638S}, we tend to rule out cases with larger coronal base densities (i.e., $n_0 \gtrsim 10^{11}~{\rm cm}^{-3}$). According to this comparison, the most plausible wind density is the one used for models 1Dip/1Map. Such a density is also able to reproduce typical emission measures of dM stars (${\rm EM} \approx 10^{51}~{\rm cm}^{-3}$) and comparatively (with the remaining cases) smaller mass-loss rates and higher wind velocities.

Because of its high mass-loss rate, V374~Peg is expected to lose about $0.01~{\rm M}_\odot$ (i.e., $\sim 3.5\%$ of its mass) during a time-scale $\tau$. If the star persists on losing this amount of mass during a longer period of time, this could have significant effects on its evolution. However, when the star ages and both its rotation as well as the stellar surface magnetic field intensity are reduced, its mass-loss rate is expected to become less intense and therefore present a less significant effect on the evolution of the star. 

There is also great interest in low-mass stars at young ages, when accretion discs are still observed. Accretion rates observed for pre-main sequence low mass stars \citep{2008ApJ...681..594H} are comparable to the wind rates that we derived here. Although it is not straightforward to extend our models to these very early phases, we note that the high wind rates we predict could have important consequences for the evolution of low-mass stars and their discs.

With the inclusion of an observed distribution of surface magnetic field, we head towards a more realistic modelling of magnetised coronal winds. Never the less, our model presents limitations, such as the neglect of a detailed energy balance. Instead, we consider a polytropic relation between pressure and density (or temperature) parametrised through $\gamma$ in the derivation of the energy equation Eq.~(\ref{eq:energy_conserve}). Once the magnetic field distribution is set, the thermal pressure adjusts itself in order to provide a distribution of heating/cooling that is able to support the MHD solution obtained \citep{1986ApJ...302..163L}. If the wind of \peg\ is able to cool down, e.g., by radiative cooling not considered in our models, the terminal velocities of the wind could be considerably smaller. Depending on where in the wind energy deposition (or removal) occurs, the wind velocity may change, without affecting the mass-loss rates. For instance, if a substantial cooling occurs above the \alf\ surface, the velocity profile of the wind from that point outwards will be affected. As the information of what is happening above the \alf\ point cannot be transmitted to the sub-\alf ic region, the wind density and velocity profiles in the proximity of the star will not be changed, and consequently neither the stellar mass-loss/angular momentum-loss rates.  

%%%%%%%%%%%%%%%%%%%%%%%%%%%%%%%%%%%%%%%%%%%%%%%%%%%%%%%%%%%%%%%%%%%%%%%%
\subsection{Departure from Potential Field}
\begin{figure}
\includegraphics[width=84mm]{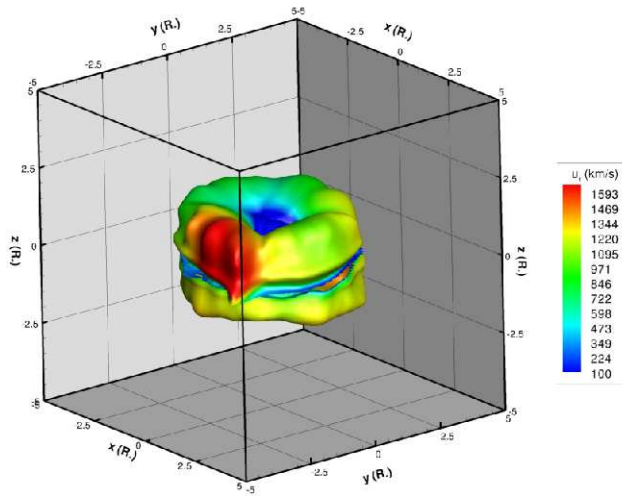}%
\caption{Isosurface of $|B_r|/B =0.97$ calculated for the MHD model (case 3Map). Color represents the radial velocity of the wind. \label{fig.ss}}
\end{figure}

Here, we compare the results of our MHD modelling of the wind of \peg\ with the output of the PFSS method, which was used as the initial configuration for the magnetic field lines in Set 2 of our simulations. Such a configuration is shown in Fig.~\ref{fig.IC}b. The spherical source surface is located at $r=r_{\rm SS}=5~R_*$ and represented by the dashed circumference. The field is considered to be everywhere potential, and its surface distribution derived from observed surface magnetic maps. Above $r=r_{\rm SS}$, the PFSS method assumes that the magnetic field is purely radial, decaying as $r^{-2}$. Ideally, we would like to compare the size of the source surface derived from the MHD modelling. However, this comparison is not straightforward, as there is no surface beyond which the magnetic field lines are purely radial in the MHD model. To overcome this difficulty, Figure~\ref{fig.ss} shows an isosurface of $|B_r|/B =0.97$, which represents the surface where $97$~percent of the magnetic field is contained in the radial component \citep{2006ApJ...653.1510R}. The first immediate conclusion from the comparison between this isosurface and the source surface of the PFSS method is the lack of sphericity of the former. Moreover, the size of the surface is considerably smaller than the radius $r_{\rm SS}=5~R_*$ adopted in the PFSS method. In our MHD model, the average size of the isosurface is $\langle r_{\rm SS} \rangle =2.6~R_*$, extending toward the poles out to $1.9~R_*$ and toward the equator out to $2.9~R_*$. We note that this size is very similar to the adopted size of $2.5$ solar radii from PFSS methods of the solar coronal magnetic field. 

The stored magnetic energy contained in the potential field is in the lowest state, i.e., it is the the minimum value of energy that the magnetic field lines can store. In the MHD wind case, excess energy is contained in the magnetic field lines due to stresses imposed by the wind. To quantify the departure of the MHD solution from the potential field solution, we evaluate the stored magnetic energy in each case. Defining the mean magnetic energy as
\begin{equation}
\langle B^2 \rangle  = \frac{\int_V B^2 {\rm d}V}{\int_V {\rm d}V},
\end{equation}
where $V$ is a given volume, we calculated the ratio $f$ between the energy contained in the MHD solution and the one in the PFSS solution
\begin{equation}
f = \frac{\langle B^2 \rangle_{\rm MHD}}{\langle B^2 \rangle_{\rm PFSS}}  .
\end{equation}
For a volume contained between $r=1$ and $5~R_*$, $f=1.5$, implying that the MHD solution has stored about $50$ percent more magnetic energy than the potential field solution. The same evaluation from  $r=1$ to $2~R_*$ results in $f=1.3$ and from $r=2$ to $5~R_*$ results in $f=2.8$. We conclude that closer to the star, the MHD solution deviates little from the potential field solution. However, the departure from a potential field becomes more important farther out from the star. 

\begin{figure}
\includegraphics[width=84mm]{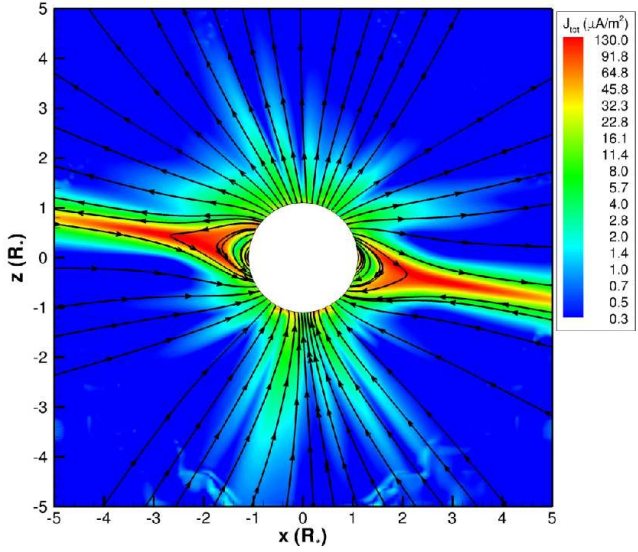}%
\caption{Current density. \label{fig.current}}
\end{figure}

The existence of currents in the MHD solution also illustrates the departure of the magnetic field configuration from a potential one. In contrast to the PFSS method, which assumes null currents everywhere (${\bf j} \propto {\bf \nabla} \times {\bf B}=0$), in the MHD wind modelling, current densities arise naturally. Figure~\ref{fig.current} shows the profile of current densities, which is similar in all the simulations of the wind of \peg. This similarity occurs because, in the ideal MHD, $j$ only depends on the intensity and configuration of the magnetic field, whose characteristics are similar in our simulations. The current density is maximum near the star and it decreases with distance. A more intense current sheet is seen separating the closed-field line region and the open-field line region. 

%%%%%%%%%%%%%%%%%%%%%%%%%%%%%%%%%%%%%%%%%%%%%%%%%%%%%%%%%%%%%%%%%%%%%%%%
\subsection{Effects on Habitability} 
The concept of habitable zones relies on the location where liquid water may be found, which is essentially determined by the luminosity of the star. Because dM stars are low-luminosity objects, the habitable zone is believed to lie near the star. However, a strong wind is able to erode the atmosphere of a planet, removing an important shield to the creation and development of life. The ram pressure $p_w$ of the wind that impacts on a planet orbiting \peg\ is
\begin{equation}
p_w = \rho u_r^2 = {\dot{m} u_r r^{-2}} ,
\end{equation}
where $\dot{m}=\rho u_r r^2$. In the asymptotic limit of the wind velocity, for large distances from the star, 
\begin{equation}\label{eq.pwinf}
p_{w, \infty} = \dot{m} u_\infty r^{-2} \simeq \frac{\dot{M} u_\infty}{4\pi r^2}.
\end{equation}
Using our scaling relations for \peg\ (cases 1Dip/1Map, 2Dip/2Map, 3Dip/3Map) $\dot{M} \simeq 4 \times 10^{-10} n_{12}^{1/2}~\msano$ and $u_\infty \simeq (1500$ -- $2300) n_{12}^{-1/2}~{\rm km~s}^{-1}$, we have
\begin{equation}
p_{w, \infty} \simeq \frac{(3 {\rm ~to~} 4.6) \times 10^{23} }{r^2} ~{\rm dyn~cm}^{-2},
\end{equation}
where the range of values exists due to different latitudinal wind velocities. We note that $p_{w, \infty}$ does not depend on the density assumed for our wind models of \peg. Therefore, even an increase in the wind mass-loss rate (e.g., through an increase in the wind base density) would not have any further effect on an orbiting planet, because there is an associated decrease in the wind velocity. 

For the solar wind
\begin{equation}
p_{w, \infty, \odot}  = \frac{\dot{M_\odot} u_{\infty,\odot}}{4\pi r^2} \simeq \frac{4 \times 10^{18} }{r^2} ~{\rm dyn~cm}^{-2},
\end{equation}
where we have used $\dot{M}_\odot = 2 \times 10^{-14}\msano$ and $u_{\infty,\odot}=400$~km~s$^{-1}$. Therefore, the wind of \peg\ has a ram pressure that is about $5$ orders of magnitude larger than for the solar wind. The immediate question we ask ourselves is how strong the planetary magnetic field should be in other to shield the planet's atmosphere from the erosive effects of the wind. Pressure balance between the wind total pressure and the planet total pressure requires that, at a distance $r_M$ from the planet,
\begin{equation}\label{eq.equilibrium}
\frac{p_w}{2} + \frac{B_w^2}{4\pi} + p= \frac{B_{{p},r_M}^2}{4\pi} + p_{p} ,
\end{equation}
where $p_w$, $B_w$ and $p$ are the local wind ram pressure, magnetic field and thermal pressure, and $B_{{p},r_M}$ and $p_p$ are the local planetary magnetic field intensity and thermal pressure. Far from the central star, $p_w \to p_{w, \infty}$, while $B_w$ and $p$ become negligible. Assuming that the planetary magnetic field is dipolar, $B_{{p},r_M} = B_p \left( {R_p}/{r_M}\right)^3$, where $B_p$ is the equatorial planetary magnetic field intensity at its surface and $R_p$ is the planetary radius. A minimum magnetic shielding of the planet requires that $r_M > R_p$. Therefore, for a planet orbiting \peg\ at $r=1~$AU, the minimum planetary magnetic field required is 
\begin{equation}\label{eq.bmin}
B_{p, {\rm min}} = (2 \pi p_{w, \infty})^{1/2} \sim 0.1~{\rm G}, 
\end{equation}
where we assumed that $p_p$ is negligible. A significant $p_p$ would only increase the limit obtained before. Therefore, for a planet orbiting far from the star ($\sim 1$~AU), a magnetic field of only $0.1$~G would be sufficient to prevent atmospheric erosion caused by the wind. In the asymptotic limit of the wind, Equations~(\ref{eq.pwinf}) and (\ref{eq.bmin}) show that $B_{p, {\rm min}} \simeq 0.1 [(1~{\rm AU})/r]$~G.

However, it should be noted that for planets orbiting closer to the star, $p_w>p_{w, \infty}$ and both the magnetic and thermal wind pressures in Equation~(\ref{eq.equilibrium}) may not be negligible. Therefore, the minimum planetary magnetic field for shielding is
\begin{equation}
B_{p, {\rm min}} = (2 \pi p_w + B_w^2 + 4\pi (p -p_p))^{1/2} .
\end{equation}
The habitable zone for a $0.5M_\odot$ star is believed to lie between $0.2$ -- $0.5$~AU, while for a $0.1M_\odot$ star, it is thought to lie much closer, between $0.02$ -- $0.05$~AU \citep{1993Icar..101..108K}. Therefore, using the results from our wind models and neglecting $p_p$, for a planet orbiting on the equatorial plane of \peg\ at $r = 0.03$~AU, 
\begin{equation}
B_{p, {\rm min}} \simeq 4~{\rm G}. 
\end{equation}
This implies that, in order to prevent atmospheric erosion, a planet orbiting inside the habitable zone of \peg\ would require a minimum magnetic field intensity of a few G. Considering that this is the magnetic field at the equator of the planet, its minimum field intensity at the pole is $\sim 8$~G, which is about half the intensity of Jupiter's magnetic field \citep[$\sim 16$~G, ][]{1998JGR...10311929C}. In fact, a planet with the same magnetic field intensity as Jupiter would have, at $0.03$~AU from \peg , a magnetospheric radius of $r_M \simeq 1.5~R_p$, where we assumed that $p_p \approx 0$.

Ultimately, when the star ages, the stellar rotation brakes, reducing the stellar surface magnetic field intensity, and therefore the wind velocity. In this case, the ram pressure felt by the planet will tend to decrease.

%%%%%%%%%%%%%%%%%%%%%%%%%%%%%%%%%%%%%%%%%%%%%%%%%%%%%%%%%%%%%%%%%%%%%%%%
\section{CONCLUSIONS}\label{sec.conclusions}
We have presented 3D MHD numerical simulations of the wind of the M dwarf star \peg. Starting from an initial magnetic field configuration and a thermally-driven wind, the wind is evolved in time, resulting in a self-consistent interaction of the wind and the magnetic field. We analysed cases where the surface distribution of magnetic field was set to be either simply dipolar or set as potential field extrapolations from observed magnetic field maps. This is the first time that an observationally derived surface magnetic field map has been implemented in a MHD model of stellar wind for a low-mass star. Our first goal was to confirm that observed maps could be reliably incorporated into the MHD code by comparing the results with those from a simple dipole. This is a necessary first step before the more complex fields of other low-mass stars \citep{2008MNRAS.390..567M, 2010MNRAS.tmp.1077M, 2008MNRAS.390..545D} can be modelled. In all our simulations, we found MHD solutions for the wind of \peg, showing that it is possible to develop coronal wind models with a realistic distribution of magnetic field. Our solutions showed that, close to the surface of the star, the magnetic field deviates little from the potential field solution. The `effective source surface' of the MHD models, i.e., where the magnetic field is mainly cointained in the radial component, has an average radius of $2.6~R_*$, about half the size of the radius adopted in the potential field solution used as initial condition.

For \peg, the model parameter that is better constrained is the surface magnetic field, from which magnetic maps have been derived. Other parameters needed by our models, such as coronal densities and temperatures, were adopted as representative of dM stars, and varied as to analyse their implications on the wind of \peg. Adopting such parameters (Table~\ref{table}), we noted that \peg\ presents a highly magnetised corona, with plasma-$\beta$ parameters of the order of $10^{-5}$ -- $10^{-3}$ at the base of the corona. This is considerably smaller than $\beta$ adopted for the base of the solar corona ($\beta_\odot \simeq 1$), which implies that the wind of \peg\ may deviate greatly from a low-velocity, low-mass-loss rate solar-type wind. In the fast magnetic rotator limit, the more simplistic stellar wind model developed by \citet{1967ApJ...148..217W} predicts terminal velocities of $\simeq 3320~{\rm km~s}^{-1}$ for a wind mass-loss rate of about $10^{-11}~\msano$ (Appendix~\ref{ap.windmodels}). 

For our models with $T_0 = 2 \times 10^6~$K and $B_0 = 1660~$G, we showed that terminal velocities are around $(1500$ -- $2300) n_{12}^{-1/2}~{\rm km~s}^{-1}$ where $n_{12} = n_0/(10^{12}~{\rm cm}^{-3}$), while the mass-loss rates are about $\dot{M} \simeq 4 \times 10^{-10} n_{12}^{1/2}~\msano$. We also estimated the characteristic time $\tau$ that the magnetised wind of \peg\ would take to remove angular momentum of the star, and therefore, brake stellar rotation. We found that the angular momentum-loss rate scales with the mass-loss rate, and that $\tau \simeq 28 {n_{12}^{-1/2}}$~Myr. Compared to observationally derived values from period distributions of stars in open clusters, this value suggests that \peg\ may have low coronal base densities $(\lesssim 10^{11}~{\rm cm}^{-3})$. Once wind densities are better observationally constrained, we will be able to further restrict our model parameters and, consequently, the derived wind characteristics. The scaling relations we found should be applicable to other highly magnetized dM stars, although their constants should differ from the ones obtained for \peg .

We investigated the possible effects the wind may have on the habitability of a planet orbiting \peg . We found that the wind ram pressure impacting on a planet is about $5$ orders of magnitude larger than the ram pressure of the solar wind for a planet orbiting at same distance. This high ram pressure of the wind of \peg\ could result in erosion of the planet's atmosphere in the case the planet does not possess a sufficiently high magnetic field able to shield it from the stellar wind. For a planet orbiting at $\sim 1$~AU, we estimated the minimum necessary magnetic field to be $\sim 0.1$~G. However, for planets much closer to the star, possibly in the habitable zone, the wind ram pressure is much larger, which could present a problem for the creation and development of life. Never the less, we estimated that a planetary magnetic field of about half Jupiter's intensity would be able to protect the planet from the erosive effects of the powerful stellar wind of \peg .% , unless the planet hosts a magnetic field that has, at least, half Jupiter's magnitude.

%%%%%%%%%%%%%%%%%%%%%%%%%%%%%%%%%%%%%%%%%%%%%%%%%%%%%%%%%%%%%%%%%%%%%%%%
\section*{Acknowledgements}
AAV acknowledges support from an STFC grant. MO acknowledges the support by National Science Foundation CAREER Grant ATM-0747654. The simulations presented here were performed at the Columbia supercomputer, at NASA Ames Research Center.

%%%%%%%%%%%%%%%%%%%%%%%%%%%%%%%%%%%%%%%%%%%%%%%%%%%%%%%%%%%%%%%%%%%%%%%%

%%%%%%%%%%%%%%%%%%%%%%%%%%%%%%%%%%%%%%%%%%%%%%%%%%%%%%%%%%%%%%%%%%%%%%%%
\appendix
\section{Comparison to Simpler Wind Models}\label{ap.windmodels}
It is interesting to have in mind the wind characteristics expected from more simplistic wind models. Non-rotating, non-magnetised, thermally-driven winds \citep{1958ApJ...128..664P} predict terminal velocities that are of the order of the Parker velocity \citep{1976ApJ...210..498B}
\begin{equation}\label{eq.velocity-parker}
u_\infty^{\rm parker} = \left( \frac{2 c_{s,0}^2}{\gamma -1} -v_{\rm esc,0}^2 + u_{r,0}^2\right)^{1/2},
\end{equation}
where $c_{s,0} = [\gamma k_B T_0/(\mu m_p)]^{1/2}$ is the sound speed, $v_{\rm esc,0} = (2GM_*/R_*)^{1/2}$ is the surface escape velocity, and $u_{r,0}$ is the radial wind velocity at $R_*$. Neglecting $u_{r,0}$ in face of the other velocities in Eq.~(\ref{eq.velocity-parker}), for \peg, $u_\infty^{\rm parker} \simeq 290~{\rm km~s}^{-1}$ for $T_0 = 2 \times 10^6~$K, and $u_\infty^{\rm parker} \simeq 1290~{\rm km~s}^{-1}$ for $T_0 = 10^7~$K. 

Because \peg\ is a highly magnetised star, spinning with a rotational period of about $\sim 11~$h, \peg\ is a fast magnetic rotator, where magneto-centrifugally force provides an efficient mechanism to accelerate the wind. On the basis of the stellar wind model developed by \citet{1967ApJ...148..217W} for a magnetised (with a radial magnetic field), rotating star, the wind terminal velocity in the fast magnetic rotator limit is given by the Michel velocity \citep{1969ApJ...158..727M}
\begin{equation}\label{eq.velocity-michel}
u_\infty^{\rm michel} = \left[ \frac{\Omega_0^2 (B_{r,0} R_*^2)^2}{\dot{M}} \right]^{1/3}.
\end{equation}
In such a limit, $u_\infty^{\rm michel} \gg c_{s,0}$ and $u_\infty^{\rm michel} \gg u_\infty^{\rm parker}$. For \peg,
\begin{equation}
u_\infty^{\rm michel} \simeq 3320 \dot{M}_{11}^{-1/3}~{\rm km~s}^{-1}, 
\end{equation}
where $\dot{M}_{11} = \dot{M}/(10^{-11}~\msano)$. For cases 1Dip/1Map, where $\dot{M}_{11} \approx 4$, $u_\infty^{\rm michel} \simeq 2100~{\rm km~s}^{-1}$, while for cases 3Dip/3Map, where $\dot{M}_{11} \approx 40$, $u_\infty^{\rm michel} \simeq 970~{\rm km~s}^{-1}$. 
It is interesting to compare these values with the ones obtained in our simulations for cases 1Dip/1Map, 2Dip/2Map, and 3Dip/3Map (i.e., $u_\infty \approx [1500$ -- $2300] n_{12}^{-1/2}~{\rm km~s}^{-1}$). Compared to the corresponding wind terminal velocity $u_\infty$, $u_\infty^{\rm michel}$ is about $7$ to $11$ times smaller than $u_\infty$ for cases 1Dip/1Map, and about $1.6$ to $2.4$ times smaller for cases 3Dip/3Map.

\bsp

\label{lastpage}
\end{document}